\documentclass[useAMS,usenatbib]{mn2e}

\usepackage{color}
\usepackage{soul}  %(for strikethrough)
\usepackage{mathrsfs} % (for script font)
\usepackage{amssymb}
\usepackage{ctable} % (for table notes)
\usepackage{epsfig}
\usepackage{epstopdf}

\usepackage[usenames,dvipsnames]{xcolor}
\usepackage{amsmath}
\usepackage{pifont}% http://ctan.org/pkg/pifont
\usepackage{multirow}
\usepackage{makecell}

\newcommand{\vc}[1]{\textit{\textbf{#1}}}  % Vector command. Can make either bf, or bf+it
\def\kms{\, {\rm km} \, {\rm s}^{-1}}
\def\mpch{\,h^{-1}\,{\rm Mpc}}
\def\lcdm{{\Lambda \textrm{CDM}}}
\def\planck{{\it Planck }}

\def\magf{{248\pm58}}
\def\mags{{243\pm58}}
\def\glf{{318\pm20}}
\def\gbf{{40\pm13}}
\def\gls{{318\pm20}}
\def\gbs{{39\pm13}}
\def\glfcirc{{318^\circ\pm20^\circ}}
\def\gbfcirc{{40^\circ\pm13^\circ}}
\def\glscirc{{318^\circ\pm20^\circ}}
\def\gbscirc{{39^\circ\pm13^\circ}}

\date{}

\title[6dFGS bulk flow]{The 6dF Galaxy Survey: Bulk Flows on $50-70 \, h^{-1} \, {\rm Mpc}$ scales}
\author[M. I. Scrimgeour et al.]
{Morag I. Scrimgeour$^{1,2,3,4}$\thanks{E:mail: morag.astro@gmail.com},
Tamara M. Davis$^5$,
Chris Blake$^6$, Lister Staveley-Smith$^{3,4}$,
\newauthor
Christina Magoulas$^{7,8,9}$,
Christopher M. Springob$^{3,4,9}$,
Florian Beutler$^{10}$,
\newauthor
Matthew Colless$^{11}$,
 Andrew Johnson$^6$,
 D. Heath Jones$^{12,13}$,
 Jun Koda$^{4,6}$, 
   \newauthor
John R. Lucey$^{14}$,
 Yin-Zhe Ma$^{15}$,
Jeremy Mould$^{4,6}$ and Gregory B. Poole$^{8}$
\\
$^1$ Department of Physics and Astronomy, University of Waterloo, Waterloo, ON, N2L 3G1, Canada \\
$^2$ Perimeter Institute for Theoretical Physics, 31 Caroline St. N., Waterloo, ON, N2L 2Y5, Canada \\
$^3$ International Centre for Radio Astronomy Research, M468, University of Western Australia, 35 Stirling Hwy, Crawley, WA 6009, Australia\\
$^4$ ARC Centre of Excellence for All-sky Astrophysics (CAASTRO) \\
$^5$ School of Mathematics and Physics, University of Queensland, Brisbane, QLD 4072, Australia\\
$^6$ Centre for Astrophysics \& Supercomputing, Swinburne University of Technology, P.O. Box 218, Hawthorn, VIC 3122, Australia \\
$^{7}$ Department of Astronomy, University of Cape Town, Private Bag X3, Rondebosch, 7701, South Africa \\
$^{8}$ School of Physics, University of Melbourne, Parkville, VIC, 3010, Australia \\
$^{9}$ Australian Astronomical Observatory, PO Box 915, North Ryde, NSW, 1670, Australia \\
$^{10}$ Lawrence Berkeley National Lab, 1 Cyclotron Rd, Berkeley CA 94720, USA \\
$^{11}$ Research School of Astronomy \& Astrophysics, Australian National University, Australia \\
$^{12}$ Department of Physics and Astronomy, Macquarie University, Sydney, NSW 
2109, Australia \\
$^{13}$ School of Physics, Monash University, Clayton, VIC 3800, Australia \\
$^{14}$ Department of Physics, University of Durham, Durham DH1 3LE, UK \\
$^{15}$ Astrophysics and Cosmology Research Unit, School of Chemistry and Physics, University of KwaZulu-Natal, Durban, South Africa \\
}

\pagerange{\pageref{firstpage}--\pageref{lastpage}} \pubyear{2002}

\def\LaTeX{L\kern-.36em\raise.3ex\hbox{a}\kern-.15em
    T\kern-.1667em\lower.7ex\hbox{E}\kern-.125emX}

\begin{document}

\maketitle

\begin{abstract}
We measure the bulk flow of the local Universe using the 6dF Galaxy Survey peculiar velocity sample (6dFGSv), the largest and most homogeneous peculiar velocity sample to date. 6dFGSv is a Fundamental Plane sample of $\sim10^4$ peculiar velocities covering the whole southern hemisphere for galactic latitude $|b| > 10^\circ$, out to redshift ${z=0.0537}$. We apply the `Minimum Variance' bulk flow weighting method, which allows us to make a robust measurement of the bulk flow on scales of $50$ and $70\mpch$. We investigate and correct for potential bias due to the lognormal velocity uncertainties, and verify our method by constructing $\lcdm$ 6dFGSv mock catalogues incorporating the survey selection function. For a hemisphere of radius $50\mpch$ we find a bulk flow amplitude of $U=\magf\kms$ in the direction $(l,b) = (\glfcirc,\gbfcirc)$, and for $70\mpch$ we find $U=\mags\kms$, in the same direction. Our measurement gives us a constraint on $\sigma_8$ of $1.01^{+1.07}_{-0.58}$. Our results are in agreement with other recent measurements of the direction of the bulk flow, and our measured amplitude is consistent with a $\lcdm$ prediction.

\end{abstract}


\color{black}

\section{Introduction}
\label{section_6dfgs_intro}

The standard model of cosmology, Lambda Cold Dark Matter ($\lcdm$) is now well supported by a wide variety of observational probes, yet questions still remain about the nature of dark matter, and whether the observed cosmic expansion is caused by a cosmological constant, $\Lambda$, or some other form of dark energy.  
Galaxy peculiar velocities are one of the only probes of large-scale structure in the nearby Universe, and are gaining interest as a promising cosmological probe that offers new information on these problems at low redshift. 
Peculiar velocities are the motions of galaxies caused by gravitational infall into local matter overdensities.  They are usually measured statistically via redshift-space distortions \citep{kaiser1987,peacock2001,tegmark2004,guzzo2008} but can also be measured directly. The line-of-sight component of the peculiar velocity $\bf{v}$ of a galaxy at position $\bf{r}$ is given by    
\begin{equation}
\label{eqn_pec_vel}
v \equiv {\bf v}\cdot \hat{\bf r} = c \left( \frac{z_{\rm obs}-z_r}{1+z_r} \right), 
\end{equation}
where $c$ is the speed of light, $z_{\rm obs}$ is the observed redshift, measured spectroscopically and corrected to the CMB restframe, and $z_r$ is the redshift corresponding to the real-space comoving distance $r$ of the galaxy.\footnote{Equation~\ref{eqn_pec_vel} is often approximated in the literature as ${v = cz_{\rm obs} - H_0 D}$, where $H_0$ is the Hubble constant and $D$ is the proper distance to the galaxy. However, this is only accurate for $z\ll0.1$ \citep{harrison1993,davis2004,davis_scrimgeour2014}.} The hat on $ \hat{\bf r}$ denotes the unit vector.

In the linear regime, the velocity field ${\bf v}({\bf r})$ is directly related to the density field $\delta({\bf r})$, via \citep{peebles1980}
\begin{equation}
\label{pv_eqn}
{\bf v}({\bf r}) = \frac{H_0 a f}{4 \pi} \int {\rm d}^3 {\bf r}' \frac{\delta({\bf r}')({\bf r}' - {\bf r})}{|{\bf r}' - {\bf r}|^3},
\end{equation}
where 
$f \equiv {\rm d} \ln D / {\rm d} \ln a$ is the present-day growth rate of cosmic structure (in terms of the linear growth factor $D$ and cosmic scale factor $a$), and $ \delta({\bf r}) = [\rho({\bf r})-\bar{\rho}]/\bar{\rho}$
with $\bar{\rho}$ the average density of the Universe. Peculiar velocity measurements therefore allow us to trace the total matter distribution, including dark matter, without the complication of galaxy bias, and over a large range of scales. They also probe the nature of gravity through the growth rate $f$.

The dipole of the velocity field, or `bulk flow' is particularly interesting since it measures the large-scale streaming motion of matter in the local Universe, which is sensitive to the large-scale modes of the matter power spectrum, and the matter density. There has been a lot of interest in the bulk flow on scales of $50 - 100 \mpch$, since some authors have suggested it is larger than expected in $\lcdm$; however, there has been a history of conflicting results in the literature. Some early measurements gave indications of apparently large bulk flows \citep{rubin1976,dressler1987b,lynden-bell1988}, while others found values consistent with predictions \citep{hart1982, devaucouleurs1984,aaronson1986} -- see \cite{kaiser1988} and \cite{strauss1995} for a review of early measurements.  More recently, an increase in the amount and quality of peculiar velocity data has led to a surge of new measurements.  Again, some of these claimed to find evidence of an unusually large bulk flow \citep{kashlinsky2008,watkins2009,feldman2010,abate2012}, while most find results consistent with $\lcdm$ \citep{colin2011,nusser2011,osborne2011,dai2011,turnbull2012,lavaux2013,ma2013,planck2013_pv,carrick2014,ma_pan2014,hong2014,feix2014}. 

Some reported detections of unusually large bulk flows have been directly challenged. \cite{kashlinsky2008} claimed to find a large dipole in the WMAP kinetic Sunyaev-Zel'dovich (kSZ) effect, indicating a bulk flow of 600-1000 km s$^{-1}$ out to $z\sim0.1$, while \cite{keisler2009} showed their uncertainties were underestimated, reducing the significance of their result.  \cite{watkins2009} combined several different peculiar velocity catalogues, and used a `minimum variance' bulk flow estimator to find a bulk flow of $407 \, {\rm km}\,{\rm s}^{-1}$ on a scale of $50\, h^{-1}\,{\rm Mpc}$, while \cite{ma2013} repeated their analysis using a hyperparameter method to combine the surveys, along with a different choice of velocity dispersion parameter, and found a smaller bulk flow consistent with $\lcdm$. Large-scale bulk flows also appear to contradict measurements of large-scale homogeneity in the galaxy distribution by \cite{hogg2005} and \cite{scrimgeour2012}. Hence, although a large bulk flow remains an intriguing possibility, it could be attributed to unaccounted-for systematic or statistical errors in existing measurements.

Another aim of measuring the large scale bulk flow is to put in context the motion of the Local Group (LG) with respect to the CMB, i.e. the bulk flow on the scale of a few Mpc. The LG motion is $627\pm22 \,{\rm km}\,{\rm s}^{-1}$ towards $l=276\pm3^\circ,b=30\pm2^\circ$ \citep{kogut1993}. In the gravitational instability model of linear theory, this is expected to be influenced by both nearby and large scale structures, and would converge to the CMB dipole when averaging over a region of sufficiently large radius.  However, attempts to reconstruct the CMB dipole using the density field have been inconsistent. 
Studies have suggested that it is necessary to go to scales of at least that of the Shapley Supercluster at $150\,h^{-1}\,{\rm Mpc}$ to recover the dipole motion \citep{kocevski2006,munoz2008,lavaux2010} while \cite{erdogdu2006a,erdogdu2006b} suggest only $\sim30\%$ of the motion is due to structures beyond $50\, h^{-1}\,{\rm Mpc}$. Other studies show no convergence up to $200-300 \mpch$ \citep{bilicki2011,nusser2014}.

In this work we aim to shine new light on the local bulk flow, using peculiar velocity data from the 6-degree Field Galaxy Survey \cite[6dFGS,][]{jones2004,magoulas2012}. This dataset is the largest, most homogeneously derived peculiar velocity sample to date, with 8885 Fundamental Plane distances. We apply the optimal Minimum Variance weighting method proposed by \cite{watkins2009,feldman2010} to measure the bulk flow.

This paper is structured as follows. In Section~\ref{section_6dfgs_peculiar_velocity_sample} we describe the 6dFGSv peculiar velocity sample.  In Section~\ref{deriving_vpec} we explain how we derive peculiar velocities from the logarithmic distances, and our method of defining the velocity uncertainty of each galaxy to avoid bias in the estimated bulk flow. In Section~\ref{section_measuring_BF} we describe the Maximum Likelihood and Minimum Variance methods that we use to estimate the bulk flow. In Section~\ref{section_6dfgs_lcdm_mock_catalogues} we describe our $\Lambda$CDM-based 6dFGSv mock catalogues. We present and discuss our results in Section~\ref{section_6dfgs_results} and conclude in Section~\ref{section_6dfgs_conclusion}.

Throughout this work we assume a flat $\Lambda$CDM cosmology with parameters from the \textit{Planck} 2013 data release, of $\Omega_{\rm m} = 0.3175$, $\Omega_\Lambda = 0.6825$, $\sigma_8=0.8344$, and $H_0 = 100 h$ km s$^{-1}$ Mpc$^{-1}$ with $h=0.67$. We only use this cosmology when converting between distance and redshift, and for comparing our bulk flow results with the $\Lambda$CDM predicted velocity dispersion. Since 6dFGSv is at low redshift ($z \le 0.054$) the results are only weakly dependent on the values of the cosmological parameters we assume. The uncertainties on these parameters are also significantly smaller than the uncertainties on our measurement, assuming a $\lcdm$ model, and so we fix these parameters throughout this work, since varying them would have a negligible effect.

%***********************************************************************************************************************************************
\section{6\lowercase{d}FGS peculiar velocity sample}
\label{section_6dfgs_peculiar_velocity_sample}

The 6-degree field Galaxy Survey (6dFGS) is a combined redshift and peculiar velocity survey of almost the whole southern hemisphere, performed using the Six-Degree Field (6dF) multi-fibre spectrograph on the UK Schmidt Telescope from May 2001 to January 2006  \citep{jones2004, jones2006, jones2009}. The survey covers galactic latitudes $|b|>10^\circ$ out to a redshift of $z \sim 0.15$. The redshift survey (6dFGSz) contains $125\, 071$ near-infrared (NIR) and optically selected spectroscopic galaxy redshifts, over $17\,000\,{\rm deg}^2$ and with a median redshift of 0.053.  Targets were selected in the $JHK$ bands from the 2MASS Extended Source Catalog \cite[2MASS XSC;][]{jarrett2000}, with secondary samples in the $b_{\rm J}$ and $r_{\rm F}$ bands.

The peculiar velocity sample, denoted 6dFGSv \citep{campbell2009,campbell2014}, is a subset of 8885 bright, early-type galaxies for which distances were derived using the Fundamental Plane (FP) relation. This sample was drawn from $\sim11\,000$ galaxies in 6dFGSz with measured FP data,  in the form of velocity dispersions and photometric scalelengths \citep{campbell2014}. The sample was selected by requiring good redshift quality ($Q = 3 - 5$), $J$-band magnitude $J<13.75$, redshifts less than $16\,500 \kms$ (or $z < 0.0537$), and velocity dispersions larger than $\sigma_0 \ge 112 \kms$, with signal-to-noise S/N$> 5 {\rm \AA}^{-1}$.

\begin{figure}
\begin{center}
\includegraphics[width=9cm]{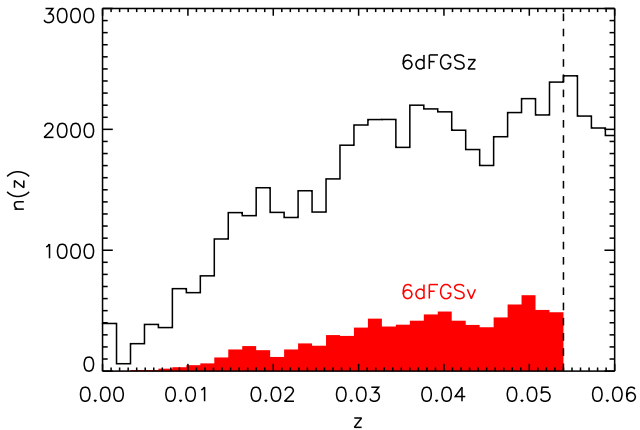}
\caption{Redshift distribution of the 6dFGS peculiar velocity sample (6dFGSv, solid red histogram) compared to the parent J-band spectroscopic sample (6dFGSz, black line histogram). The vertical dashed line shows the redshift cut imposed on the velocity sample.}
\label{6dFGSv_nz}
\end{center}
\end{figure}

The redshift distribution of 6dFGSv compared to that of the parent  J-band 6dFGSz sample is shown in Figure~\ref{6dFGSv_nz}. The fitting of the FP, and the selection cuts applied to obtain the FP and peculiar velocity samples, are described in detail in \cite{magoulas2012} (hereafter M12). The derivation of the FP distances and peculiar velocities, 
 and correction for Malmquist bias and other selection effects, is described in \cite{springob2014} (hereafter S14).
 
%----------------------------------------------------------
\section{Deriving peculiar velocities for 6\lowercase{d}FGS\lowercase{v}}
\label{deriving_vpec}

The output of the FP peculiar velocity derivation 
for 6dFGSv (from S14) is a probability distribution for the `logarithmic distance ratio' for each galaxy, $\eta$, defined by 
\begin{equation}
\label{x}
\eta \equiv \log_{10} (D_z / D_r),
\end{equation}
where $D_z$ is the co-moving distance in the fiducial $\Lambda$CDM cosmology corresponding to the observed redshift $z$, while $D_r$ is the co-moving distance corresponding to the angular diameter distance inferred from the Fundamental Plane. 

Instead of obtaining $\eta$ as a single value with an uncertainty, S14 derive the full posterior probability distributions $P(\eta)$, in order to retain all the available information resulting from the selection cuts on the FP. These probability distributions  are close to Gaussian in log distance, with a small skew due to the different selection effects and bias corrections, as described in S14.

The optimal `Minimum Variance' estimator we wish to use for the bulk flow measurement, described in the next section, takes as input peculiar velocities in $\kms$.
To convert $\eta$ to peculiar velocity $v$, we use the fact that
\begin{equation}
\label{eqn_z_pv}
(1+z) = (1+z_r)(1+z_{\rm p})
\end{equation}
where $z_r$ is the redshift corresponding to $D_r$ in the assumed cosmology, and $z_{\rm p}$ is the `peculiar redshift,' $z_{\rm p} = v/c$, where $v$ is the line-of-sight component of the galaxy's peculiar velocity. The relation between redshift and co-moving distance is
\begin{equation}
\label{comoving_dist}
D(z) = \frac{c}{H_0} \int_0^z \frac{{\rm d}z'}{E(z')}
\end{equation}
where
\begin{equation}
E(z) = \frac{H(z)}{H_0} = [\Omega_{\rm m}(1+z)^3 + \Omega_{\Lambda}]^{1/2},
\end{equation}
for which we use the fiducial $\Lambda$CDM parameter values listed in Section~\ref{section_6dfgs_intro}.

\begin{figure}
\begin{center}
\includegraphics[width=9cm]{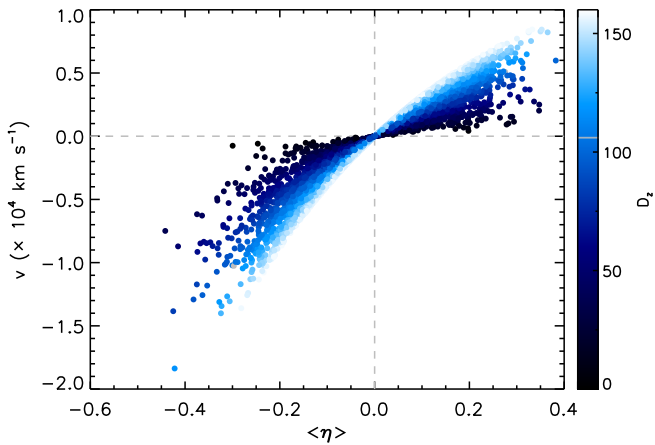}
\caption{Radial peculiar velocity $v$ from Equation~\ref{vp_x}, as a function of the mean $\eta$ value of each galaxy, ${\langle \eta \rangle}$, for the 6dFGSv sample, colour-coded by redshift distance $D_z$. The $v(\eta)$ relation is single-valued and monotonic for a given $D_z$, and is increasingly nonlinear for increasing $D_z$.}
\label{v_x}
\end{center}
\end{figure}

The peculiar velocity $v$ corresponding to $\eta$ is then
\begin{equation}
\label{vp_x}
v(\eta,z) = c \left( \frac{z-z_r(\eta,z)}{1+z_r(\eta,z)}  \right),
\end{equation}
with $z_r$ obtained from $\eta$ and $z$ using Eqs.~\ref{x} and \ref{comoving_dist}. This relation for $v(\eta,z)$ is  illustrated for the 6dFGSv sample in Figure~\ref{v_x}.

We see that $v(\eta,z)$ is nonlinear at fixed redshift, which poses a problem for obtaining an unbiased estimate of $v$. The observable quantity $\eta$ has Gaussian uncertainty in log-space, which translates to lognormal uncertainty on $v$. 
This is a standard problem for peculiar velocity measurements. 

\begin{figure}
\begin{center}
\includegraphics[width=9cm]{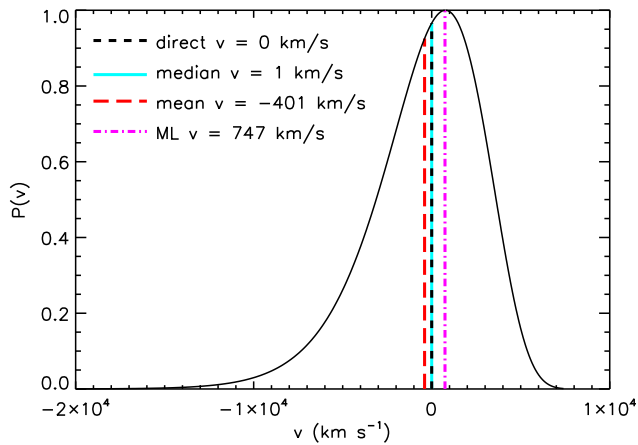}
\caption{A typical $P(v)$ distribution for 6dFGSv, for an imagined galaxy at the mean redshift of 6dFGSv, and having the mean $\eta$ uncertainty, $\sigma_\eta$, of the sample, but with $\langle \eta \rangle \equiv 0$. 
The red long-dashed line is the mean of $P(v)$, the magenta dot-dashed line is the maximum likelihood, the cyan solid line is the median and the black short-dashed line is the direct conversion of $\langle \eta\rangle$ to $v$, which is almost identical to the median.
Since $\langle \eta \rangle$ is zero, the peculiar velocity of the galaxy should be zero, but only the median and the direct $\langle \eta\rangle \to v$ have this value.} 
\label{Pv_x0_new}
\end{center}
\end{figure}

We can see this by converting the $P(\eta)$ distributions to probability distributions of velocity, $P(v)$, using the relation
\begin{equation}
\label{Pvp_eqn}
P(v) = P(\eta)\frac{{\rm d}\eta}{{\rm d}v} = P(\eta) \frac{1}{D_r\ln(10)} \frac{{\rm d} D_r}{{\rm d}z_r} \frac{(1+z_r)^2}{c(1+z)},
\end{equation}
where $\ln$ is the natural logarithm. A typical velocity probability distribution $P(v)$ is illustrated in Figure~\ref{Pv_x0_new}, where we have set $\langle \eta \rangle \equiv 0$. The distribution is close to lognormal.

We obtain the peculiar velocity for each galaxy by taking the mean value of the $P(\eta)$ distribution, $\langle \eta \rangle = \int_{-\infty}^{\infty} \eta P(\eta) {\rm d}\eta$, and converting it to velocity using Eq.~\ref{vp_x}. This is equivalent to the median of the $P(v)$ distribution, as can be seen in Fig.~\ref{Pv_x0_new}. This is the least biased way to determine velocity, correctly giving zero $v$ for zero $\eta$, and it is also the standard method used in the literature.

\begin{figure}
\begin{center}
\includegraphics[width=9cm]{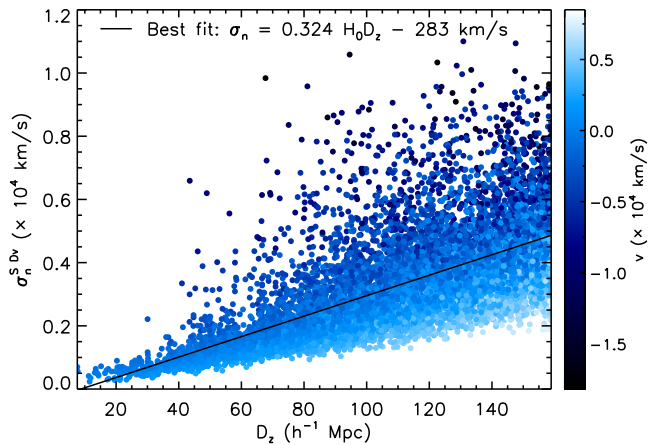}
\caption{The correlation between the standard deviation $\sigma_n^{\rm SDv}$ of $P(v)$, and redshift distance $D_z$. The colour gradient shows the corresponding peculiar velocity. A linear best-fit to the points is shown in black.} 
\label{sigmav_z3}
\end{center}
\end{figure}

Since the uncertainty on $\eta$ is Gaussian, the uncertainty on $v$ is lognormal, and is proportional to $D_z$. The uncertainty on $v$, as derived from $P(v)$, is also proportional to $v$, since the width of $P(v)$ increases with radial velocity. To derive the velocity uncertainties as approximate Gaussian uncertainties, $\sigma_v$, which we need for the Minimum Variance estimator, we first calculate the standard deviation of each $P(v)$ distribution,
\begin{equation}
\label{eqn_stddev_v}
\sigma_n^{\rm SDv} = \left( \int_{-\infty}^{\infty} v^2 P(v) {\rm d}v - \bar{v}^2 \right)^{1/2},
\end{equation}
and plot this against $D_z$, as shown in Figure~\ref{sigmav_z3}. Overall this is a linear relation, but there is a strong dependence on $v$, creating large scatter. By taking the linear best fit we remove this dependence, essentially taking the uncertainty a galaxy would have if it had zero peculiar velocity. For our bulk flow estimation, we use the velocity uncertainty given by this linear best-fit:
\begin{equation}
\sigma_{n} = 0.324\, H_0 D_z. 
\end{equation}
This approximation removes the dependence of the velocity uncertainty on the measured $v$. This is an important correction, because the weights assigned to each galaxy in the bulk flow estimation are derived depending on the galaxy's velocity uncertainty.  
If the weights were correlated with the velocities themselves, this would produce a biased bulk flow measurement, made worse if the redshift distribution of galaxies is not evenly distributed over the sky, as in 6dFGS.

%***********************************************************************************************************************************************
\section{Bulk flow estimators}
\label{section_measuring_BF}

The bulk flow is the average peculiar velocity in a given volume of space, usually taken to be a spherical region centred on us, and defined by 
\begin{equation}
{\bf U}(R) = \frac{3}{4 \pi R^3} \int_{x=0}^R {\bf v}({\bf x}) {\rm d}^3 x,
\end{equation}
where $R$ is the radius of the sphere in which the bulk flow is measured.
In practice however, we can never perfectly sample the velocity field. Peculiar velocity samples are typically sparse, with complicated geometries and large measurement uncertainties. Additionally, we only observe the line-of-sight component of the peculiar velocities. 

Different bulk flow estimators have been suggested in the literature to account for this, including the maximum likelihood estimate \citep{dressler1987,kaiser1988}, comparison with the density field \citep{bertschinger1990,dekel1999,willick1998,turnbull2012,magoulas2015}, reconstruction of the velocity field based on a velocity power spectrum \citep{nusser2011} and the so-called `Minimum Variance' weighting method \citep{watkins2009,feldman2010}.

In this paper we apply the maximum likelihood estimate (MLE) and the Minimum Variance (MV) method to 6dFGSv.
These both evaluate the bulk flow as a weighted sum of the peculiar velocities. Given a sample of objects with radial peculiar velocities $v_n$, these methods assign a weight $w_{i,n}$ corresponding to the $i^{\rm th}$ direction for each galaxy. The bulk flow $\vc{U} = (u_x,u_y,u_z)$ is then
\begin{equation}
\label{eqn_BF_estimate}
u_i = \sum_n w_{i,n} v_n.
\end{equation}
In the following two subsections we give an overview of these two weighting methods.

A parallel analysis of the 6dFGSv bulk flow is currently being made by Magoulas et al. (in prep), who apply a different bulk flow estimation method. They use forward modelling, performing
a Maximum Likelihood fit to a bulk flow model transformed into the
observational space of the Fundamental Plane parameters. This approach
effectively fits the measured logarithmic distance ratios $\eta = \log (D_z/D_r)$
without converting to linear velocities, and can fully account for the
(Gaussian) error distribution in the observational space.

%----------------------------------------------------------
\subsection{Maximum likelihood estimate}

The MLE has traditionally been the most common technique used to measure the bulk flow. We consider here the MLE using inverse variance weighting from \cite{kaiser1988}. Given a sample of $N$ objects at positions $r_{n,i}$, each having a measured line-of-sight velocity $v_n$ with uncertainty $\sigma_n$, 
the MLE weight for the $n^{\rm th}$ galaxy is
\begin{equation}
\label{eqn_MLE_weight}
w_{i,n} = \sum_j A_{ij}^{-1} \frac{\hat{r}_{n,j}}{\sigma_n^2+\sigma_*^2}
\end{equation}
where
\begin{equation}
\label{eqn_Aij}
A_{ij} = \sum_n \frac{ \hat{r}_{n,i} \hat{r}_{n,j}}{ \sigma_n^2 + \sigma_*^2  }.
\end{equation}
The parameter $\sigma_*$ is the 1D velocity dispersion, usually assumed to be $\sim 300\kms$;  we assume galaxies have random motions drawn from a Gaussian distribution with this dispersion, in addition to the bulk flow component. These random motions add to the noise for any given galaxy.

This solution makes a number of simplifying assumptions:
\begin{enumerate}
\item the observational errors, $\sigma_n$, are Gaussian
\item linear theory holds, so $v_n \ll H_0r_n$
\item we can neglect uncertainty in $r_n$
\item $u_i$ is fairly insensitive to small-scale velocities, and that $\sigma_*$, which will be strongly influenced by nonlinear flows, can be fixed at a given value. 
\end{enumerate}
In practice, nearly all of these assumptions will be violated to some extent. The observational errors on $v$ are \textit{not} Gaussian, $v \sim 20-30\%$ of $H_0r$, and linear theory does not strictly apply, since $\sigma_*\sim300\,{\rm km}\,{\rm s}^{-1}$ is comparable to the expected bulk flow amplitude on the scales we measure. However, we do not expect these to have a significant impact on our measurement. Since our analysis is done in redshift-space, the uncertainty on $r_n$ is the uncertainty on the redshift distance, which is indeed negligible.  We find that our result is insensitive to the choice of $\sigma_*$, which we discuss further in Section~\ref{section_sigstar}. We leave further analysis of non-Gaussian uncertainties to future work.

%----------------------------------------------------------
\subsection{Minimum Variance method}

Although the MLE is simple to perform,  it has several disadvantages. It will have a complex window function dependent on the geometry and uncertainties of a particular survey, making it difficult to compare between surveys and with theory. It is also density-weighted rather than volume-weighted, as it tends to upweight high-density regions where galaxies are more likely to be measured, and down-weight low density regions. Finally, because it down-weights more distant galaxies which have larger uncertainties, the MLE tends to be dominated by the nearest galaxies in the sample and so minimises the scale on which the bulk flow is measured.

The `Minimum Variance' (MV) method of \cite{watkins2009} (hereafter WFH09) and \cite{feldman2010} is an extension of the MLE method, which constructs a more optimal set of weights that allow a volume-weighted measurement of the bulk flow to be made with a specified window function. This is achieved by determining weights $w_{i,n}$  that minimise the variance between the bulk flow measured by the sample, and the bulk flow that would be measured by an `ideal' survey, with the specified window function. In their case, they choose this to be a perfectly sampled, all-sky Gaussian survey with `ideal' radius $R_I$. 

While the MV method is more optimal than the MLE method, it still has some disadvantages. It is not necessarily an unbiased estimator, especially since it still assumes the velocity uncertainties are Gaussian, and it minimises the variance only on the particular quantity it tries to measure (i.e. the bulk flow of a given window function) rather than the bulk flow of the full dataset. However, it provides a much more optimal way of comparing the bulk flow in a survey with a theoretical model and with other surveys. Tests of its robustness using $N$-body simulations have shown that it correctly recovers the underlying bulk flow, and is unbiased by the survey geometry and nonlinear flows \citep{agarwal2012}.

The MV weights are calculated from
\begin{equation}
\label{eqn_MV_weight_1}
\mathbf{w}_i = (\mathsf{\bf G}+\lambda \mathsf{\bf P})^{-1}{\bf Q}_i,
\end{equation} 
where $i$ denotes the three bulk flow components. $\mathsf{\bf P}$ is the $k=0$ limit of the angle-averaged window function, ${\bf Q}_i$ incorporates information about the input ideal window function,  $\lambda$ is a Lagrange multiplier, and $\mathsf{\bf G}$ is the covariance matrix of the individual peculiar velocities, given by
\begin{eqnarray}
\label{eqn_Gnm_1}
G_{nm} &=& \langle v_n v_m \rangle \\
&=& \delta_{nm}(\sigma_n^2 + \sigma_*^2) + \frac{(f(\Omega_{\rm m},z)H_0a)^2}{2\pi^2} \int P(k)f_{mn}(k){\rm d}k, \nonumber
\end{eqnarray}
where $H_0$ is the Hubble constant, $f\sim\Omega_{\rm m}^{0.55}(z)$ is the growth rate of cosmic structure, and $f_{mn}(k)$ is the angle-averaged window function,
\begin{equation}
\label{eqn_fmnk_1}
f_{mn}(k) = \int \frac{{\rm d}^2\hat{k}}{4 \pi} (\hat{\textbf{\textit{r}}}_n \cdot \hat{\textbf{\textit{k}} })(  \hat{\textbf{\textit{r}}}_m \cdot  \hat{\textbf{\textit{k}}} ) \times \exp [ik\hat{\textbf{\textit{k}}} \cdot ( \textbf{\textit{r}}_n-\textbf{\textit{r}}_m)].
\end{equation}
The first term in Equation~\ref{eqn_Gnm_1} is the noise term, while the second part is the cosmic variance, or `geometrical' term, and incorporates the power spectrum of a given cosmological model. Equation~\ref{eqn_fmnk_1} can be calculated analytically, as shown in the appendix of \cite{ma2011}. Further details of how the weights are calculated are presented in Appendix~\ref{MV_appendix}; also see WFH09 and \cite{feldman2010}.

Following WFH09 we also choose a Gaussian survey as our ideal survey, using two different ideal radii: (1) $R_I=50\,h^{-1}\,{\rm Mpc}$ for comparison with WFH09; and (2) $R_I=70\,h^{-1}\,{\rm Mpc}$. We choose the latter since it is close to the `Maximum Likelihood Estimate depth' of 6dFGSv, which is calculated via
\begin{equation}
d_{\rm MLE} = \frac{\sum r_n w_n}{\sum w_n},
\end{equation}
where the MLE weights are $w_n = 1/(\sigma_n^2 + \sigma_*^2)$. We find this to be $\sim 72 \,h^{-1}\,{\rm Mpc}$ for 6dFGSv. This is the optimal depth for a bulk flow measurement in 6dFGSv.

\begin{figure}
\begin{center}
\includegraphics[width=9cm]{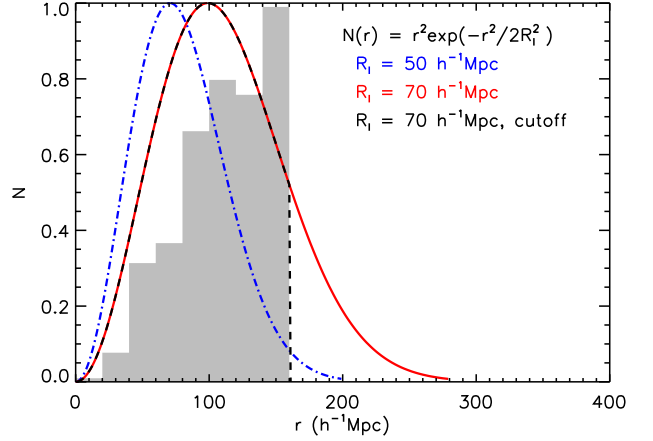}
\caption{Radial number distribution $N(r)$ of the Gaussian filters we use for measuring the 6dFGSv bulk flow.  The filters have (i) $R_I = 50 h^{-1}$Mpc radius (blue dot-dashed line), (ii) $R_I = 70 h^{-1}$Mpc radius (red solid line) and (iii) $R_I = 70 h^{-1}$Mpc radius with a cutoff at $160h^{-1}$Mpc, corresponding to the redshift cut of the data (black dashed). The distribution of the data is shown by the grey histogram for comparison. }
\label{Gaussian_radialdist_6dFGS}
\end{center}
\end{figure}

The ideal Gaussian survey will have a radial density profile given by
\begin{equation}
\rho( r ) \propto \exp(-r^2/2R_I^2),
\end{equation}
and its radial number distribution is
\begin{equation}
N( r ) \propto r^2 \exp(-r^2/2R_I^2).
\end{equation}
We plot $N(r)$ for our two ideal surveys in Figure~\ref{Gaussian_radialdist_6dFGS}, along with the number distribution of 6dFGSv for comparison. The 6dFGSv sample has a cutoff at $160\,h^{-1}\,{\rm Mpc}$ corresponding to $z=0.0537$, so we also apply this to our ideal surveys.

The ideal survey used by WFH09 is an all-sky survey, since the dataset they used was all-sky; in the case of 6dFGSv, we only have half the sky. We discuss the effect of partial sky coverage on our measurement in Section~\ref{section_6dfgs_results}.

\subsection{Bulk flow uncertainties}
\label{section_BF_uncertainties}

The covariance matrix of the bulk flow moments, $R_{ij}$,  can be written as:
\begin{equation}
\label{eqn_Rij_full}
R_{ij} = \langle u_i u_j \rangle = \sum_{mn} w_{im}w_{jn} G_{mn} = R_{ij}^{(\epsilon)} + R_{ij}^{(v)},
\end{equation}
where $R_{ij}^{(\epsilon)}$ represents the noise contribution,
\begin{equation}
R_{ij}^{(\epsilon)} = \sum_n w_{i,n}w_{j,n}(\sigma_n^2+\sigma_*^2),
\end{equation}
and  $R_{ij}^{(v)}$ represents the cosmic variance contribution,
\begin{equation}
R_{ij}^{(v)} = \frac{(f(\Omega_{\rm m},z)H_0a)^2}{2\pi^2} \int_0^\infty {\rm d}k \mathcal{W}^2_{ij}(k) P(k),
\end{equation}
where $\mathcal{W}^2_{ij}(k)$ is the angle-averaged tensor window function,
\begin{equation}
\label{eqn_tensorwinfun}
\mathcal{W}^2_{ij}(k) =  \sum_{n,m} w_{i,n} w_{j,m} f_{mn}(k).
\end{equation}
The errors on the bulk flow moments, $\sigma_i$, are then ${\sigma_i = \sqrt{R_{ii}}}$, and the error on the bulk flow magnitude is $\sigma_U^2 = J\, R_{ij} J^{\rm T}$, where $J$ is the Jacobian of $U$, $\partial U / \partial u_i$. 

%----------------------------------------------------------
\subsection{$\sigma_*$ estimation}
\label{section_sigstar}

The 1D velocity dispersion parameter $\sigma_*$, as previously mentioned, accounts for small-scale random motions. The value of $\sigma_*$ affects the weights of nearby galaxies most strongly, since they have the smallest velocity errors, but in the Minimum Variance method where these are down-weighted by the ideal window function,  $\sigma_*$ will only have a small effect on the measured bulk flow \citep{feldman2010}.  

In the case of 6dFGSv, small-scale velocities need to be accounted for in the fitting of the FP, since the fitting is done assuming each galaxy is at its redshift distance, and so velocities add to the scatter of the plane.  A value of $\sigma_*=300\,{\rm km}\,{\rm s}^{-1}$ is accounted for in the fitting of the FP by M12 and S14, and is effectively subtracted from the uncertainty in the $P(\eta)$ distributions.  This means we need to `add back in' this uncertainty in our bulk flow weights. \cite{johnson2014} perform a fit to $\sigma_*$  for the 6dFGSv sample and find it to peak at zero. However, $\sigma_*$ also acts to regularise the bulk flow weights, to prevent galaxies with low error dominating the results, so assuming a zero $\sigma_*$ is not ideal. We find that varying $\sigma_*$ from $0$ to $250\kms$ has little effect on our results, changing the MV bulk flow on the order of $\sim2\%$. We therefore fix $\sigma_*= 250 \kms$ for our analysis.

%***********************************************************************************************************************************************

\section{6\lowercase{d}FGS\lowercase{v} Selection Function and $\Lambda$CDM Mock Catalogues}
\label{section_6dfgs_lcdm_mock_catalogues}

In order to test possible systematics in our bulk flow measurement arising from the survey selection function, we apply our bulk flow analysis to $\Lambda$CDM mock catalogues of 6dFGSv, incorporating the survey selection function. To create $\Lambda$CDM peculiar velocity mocks, we need to make use of an $N$-body simulation which provides both the positions and velocities of galaxies. In this section we describe how we determine the selection function of 6dFGSv, and use this to generate mock catalogues using the GiggleZ $N$-body simulation. This selection function also allows for the creation of random catalogues for clustering analysis. 

%----------------------------------------------------------
\subsection{6dFGSv survey selection function}
\label{section_6dfgs_selectionfunction}

The selection function $W({\bf x})$ is a function indicating the expected number density of 6dFGSv galaxies at a position $\bf x$, due to the different selection criteria of the sample. These can be both angular- and redshift- dependent. To implement the selection function in our mocks, we reproduce the selection process described in M12 and S14 to obtain the 6dFGSv sample of 8,885 galaxies from the full 6dFGS redshift sample of ~125,000 galaxies.  In summary, \color{black} they first select galaxies suitable for fitting the FP, by choosing galaxies with reliable redshifts (with redshift quality $Q=3-5$) and redshifts less than $16\,500\,{\rm km}\,{\rm s}^{-1}$ (or $z<0.0537$), above which a key spectral feature used to measure velocity dispersion is shifted out of the wavelength range. They then morphologically select early-type (E/S0) galaxies, by matching the observed spectra to template galaxy spectra. This produced a sample of $\sim20\,000$ galaxies.

These $\sim 20\,000$ galaxies then had their velocity dispersions measured using the Fourier cross-correlation technique \citep{campbell2009}. Of these, galaxies with a signal-to-noise S/N$> 5 {\rm \AA}^{-1}$, and velocity dispersions larger than the instrumental resolution limit ($s \ge 2.05$, or $\sigma_0 \ge 112$ km s$^{-1}$) were selected, to produce a `Fundamental Plane sample' of $11\,287$ galaxies. This sample, with both spectroscopic measurements from 6dFGS and photometric measurements from 2MASS in the $J$, $H$ and $K$ bands, was used by M12 for the fitting of the FP parameters.

Finally, the peculiar velocity sample 6dFGSv was obtained from the FP sample after several further cuts. A stricter redshift limit of $cz < 16120$ ($z<0.0537 $) was imposed in the CMB frame, along with further magnitude cuts of $J\le 13.65,H\le12.85$ and $K\le12.55$, to maintain high completeness over the sky. Further galaxies were removed after a visual inspection, and a velocity dispersion $\chi^2$ cut, to obtain the final peculiar velocity sample of 8885 galaxies.

%----------------------------------------------------------
\subsection{Fundamental Plane fitting}

Here we introduce the FP terminology we will use in making the mocks - see M12 for further details.
The FP relation can be written in logarithmic units as
\begin{equation}
r = a s + b i + c,
\end{equation}
where ${r \equiv \log R_e}$,  $s \equiv \log \sigma_0$ and  $i \equiv  \log \langle I_e \rangle$, where $R_e$ is the effective radius in units of  $h^{-1}\,{\rm kpc}$, $\sigma_0$ is the central velocity dispersion in units of ${\rm km}\,{\rm s}^{-1}$, and $\langle I_e \rangle$ is the mean surface brightness, in units of ${\rm L}_\odot \,{\rm pc}^{-2}$. The coefficients $a$ and $b$ are the slopes of the plane and $c$ is the offset of the plane. M12 use logarithms of base 10.

M12 determine the FP parameters for 6dFGSv using a maximum-likelihood fit to a 3D Gaussian model. The FP can be described either in terms of the observational parameters $(r, s, i)$, or in terms of the three unit vectors corresponding to the axes of the 3D Gaussian describing the galaxy distribution. M12 refer to these as `FP-space' and `{\bf v}-space' respectively.  The model can then be described by eight parameters: $\{a,b,\bar{r},\bar{s},\bar{i},\sigma_1,\sigma_2,\sigma_3\}$, where $(\bar{r},\bar{s},\bar{i})$ define the centre of the 3D Gaussian in FP-space and $(\sigma_1,\sigma_2,\sigma_3)$ are the dispersion of the Gaussian along each of the three axes in $\textbf{\textit{v}}$-space. The offset of the FP can be calculated as $c = \bar{r} - a\bar{s} -b\bar{i}$. 

%----------------------------------------------------------
\subsection{Mock sample algorithm}
\label{mocksamplealgorithm}

We create mock $\Lambda$CDM realisations of the 6dFGSv dataset for the set of FP parameters $\{ a,b,c, \bar{r}, \bar{s}, \bar{i}, \sigma_1, \sigma_2, \sigma_3 \}$ derived by M12. 
We use the following steps to reproduce the 6dFGSv selection function and generate the mock catalogue: 
\begin{enumerate}
\newcounter{saveenum}
\item For a $\lcdm$  mock, start by drawing haloes from an $N$-body simulation in a mass range equivalent to the 6dFGS elliptical galaxies, i.e. pick haloes that match the bias of 6dFGS (this is effectively a cut in morphological type).
\setcounter{saveenum}{\value{enumi}}
\end{enumerate}

\textbf{Angular \& Redshift cuts}
\begin{enumerate}
\setcounter{enumi}{\value{saveenum}}
\item Define the location of the observer, and calculate RA, Dec, true comoving distance $D_r$ and radial peculiar velocity $v$ for each galaxy. Also calculate the true and observed redshifts $z_r$, $z$, using Equation~\ref{eqn_z_pv}. 
\item Only include haloes within hard angular cuts ${{\rm Dec} <  0^\circ}$ and Galactic latitude $|b|>10^\circ$.
\item Impose a redshift cut of $cz < 16\,120$ km s$^{-1}$. 
\item Normalise by applying a random subsampling to obtain the number of galaxies in the 6dFGS parent redshift sample.
\setcounter{saveenum}{\value{enumi}}
\end{enumerate}
\textbf{Magnitude \& Velocity Dispersion cuts}
\begin{enumerate}
\setcounter{enumi}{\value{saveenum}}
\item For each galaxy, draw values for $v_1$, $v_2$ and $v_3$ at random from a 3D Gaussian with standard deviations $\sigma_1$, $\sigma_2$ and $\sigma_3$ as listed in Table 3 of M12. We use the J-band values as the J-band has the smallest photometric errors.
\item Transform these values from the $\textbf{\textit{v}}$-space (principle axes) coordinate system to the $\{ \textbf{\textit{r}}, \textbf{\textit{s}}, \textbf{\textit{i}}  \}$-space (observed parameters) coordinate system using the inverse of Equation~6 in M12,
with the specified FP slopes ($a$ and $b$) and FP mean values ($\bar{r}$, $\bar{s}$ and $\bar{i}$). This gives the true Fundamental Plane parameters $(r_t,s_t,i_t)$ for the simulated galaxies.
\item Re-order each set of $(r_t,s_t,i_t)$ parameters in descending order of luminosity,
\begin{equation}
\log L = l = 2r + i,
\end{equation}
and assign them to the haloes in descending order of maximum circular velocity $V_{\rm max,sub}$.

\item Use the comoving distance $D_r$ of each galaxy from the observer to determine the angular radius $\theta$ from the physical radius $r_t$, by calculating the angular diameter distance $D_{\rm A}$:
	\begin{equation}
	D_{\rm A} \equiv \frac{r_t}{\theta }  = \frac{D_r}{1+z_{\rm true}}
	\end{equation}
	\citep[this relation is true for $\Omega_k=0$, see][]{hogg1999}.
	Then $\theta$ is obtained from
	\begin{equation}
		\log \theta = \log r_t - \log D_{\rm A}.
		\end{equation}	
\item Determine the true apparent magnitude $m_t$ from the angular radius $\theta$ and the \textit{degraded} surface brightness $i$ using
\begin{equation}
m_t = \langle \mu_e \rangle - 2.5 \log [2\pi\theta^2],
\end{equation}	
where $\langle \mu_e \rangle = M - 2.5i + 21.57$, where $M = 3.67$ for the $J$-band. The surface brightness $i$ is first degraded by `de-correcting' for K-correction and surface brightness dimming.
\item 
Obtain the correlated measurement uncertainties in $r$, $s$ and $i$, $(\epsilon_r,\epsilon_s,\epsilon_i)$, from the magnitude $m_t$, using the matrix in Equation~13 of M12. 
\item Add these measurement errors to $\{ r, s, i \}$ to obtain the observed values $\{ r_o,s_o,i_o \}$ for each galaxy.
\color{black}
\item Only include galaxies with velocity dispersion ${s_o>\log(116 {\rm \,km/s})}$. (Cut for instrumental resolution).
\item Determine the observed magnitude $m_o$ using the observed values $r_o$ and $i_o$. 
\item Keep the galaxy if the observed magnitude $m_o$ is brighter than the faint limit for the velocity sample (${J \le 13.65}$). 
\item Use the selection function described in \cite{jones2006} to determine the angular completeness of the 6dFGS spectroscopic follow-up, given the (RA, Dec, $m_o$) values for each galaxy. Sub-sample the galaxies with this probability.
\item Apply a random subsampling to account for cuts in signal-to-noise (S/N) ratio and $R$.
\end{enumerate}

\subsection{The Mocks}
In order to generate $\Lambda$CDM mocks, we apply our mock sample algorithm to the GiggleZ (Giga-parsec WiggleZ) simulation. GiggleZ \citep{poole2014} is a suite of dark matter $N$-body simulations run at Swinburne University of Technology. It has a WMAP-5 cosmology with $(\Omega_\Lambda,\Omega_{\rm m},\Omega_b,h,\sigma_8,n) = (0.727, 0.273, 0.0456, 0.705, 0.812, 0.960)$. 
We use the GiggleZ Main simulation, which contains $2160^3$ dark matter particles in a periodic box of side $1 h^{-1}$Gpc. The particle mass is $7.5 \times 10^9 h^{-1}M_\odot$, which allows bound systems with  masses $\gtrsim 1.5 \times 10^{11} h^{-1} M_\odot$ to be resolved. 

Halo finding for GiggleZ was performed using \textsc{subfind} \citep{springel2001}, which utilises a friends-of-friends (FoF) algorithm to identify coherent overdensities of particles and a substructure analysis to determine bound overdensities within each FoF halo. We place a galaxy at the centre of each subhalo, and rank-order them by their maximum circular velocity ($V_{\rm max,sub}$) to obtain the largest haloes, in order to reproduce the bias of the 6dFGSv sample. 

\begin{figure}
\begin{center}
\includegraphics[width=9cm]{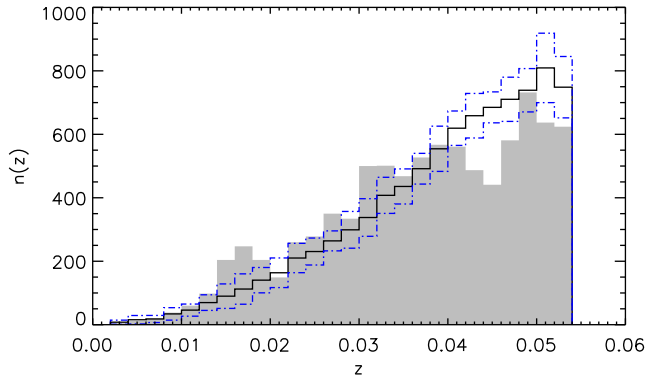}
\caption{The mean redshift distribution of our 20 GiggleZ mocks (black histogram), along with the standard deviation (blue dot-dashed histograms), compared to 6dFGSv (solid grey histogram).} 
\label{nz_mean_vmock_Gz_highbias_c0p6}
\end{center}
\end{figure}

We have generated 20 independent mocks of 6dFGSv  within the GiggleZ volume. We show the mean and variance of the redshift distribution of our 20 mocks, compared with 6dFGSv, in Figure~\ref{nz_mean_vmock_Gz_highbias_c0p6}. The mocks appear higher than the data in the highest redshift bins $(0.04 < z < 0.05)$, although this could possibly be attributed to cosmic variance. However, the large-scale bulk flow properties of the mocks will not depend strongly on the exact shape of the redshift distributions, since the bulk flow depends to first order on the velocities of galaxies, not on their number density. 

%***************************************************************************************************
\section{Results and Discussion}
\label{section_6dfgs_results}

We present in this section the bulk flow results of our 6dFGSv analysis, for the MV and MLE estimators, along with the bulk flow results for our $\Lambda$CDM mocks. We then compare our results to a theoretical $\Lambda$CDM prediction, firstly considering the 3D bulk flow amplitude, and secondly considering each of the three 1D bulk flow components, to obtain constraints on $\Omega_{\rm m}$ and $\sigma_8$. 

%----------------------------------------------------------
\subsection{Bulk flow results}
\label{section_bf_results}

We have calculated the bulk flow for 6dFGSv, for the two different bulk flow estimators described in Section~\ref{section_measuring_BF}:
\begin{enumerate}
\item The Minimum Variance (MV) estimate, using two different ideal surveys: (1) a Gaussian survey with effective radius $R_I = 50\, h^{-1}\,{\rm Mpc}$, (2) a Gaussian survey of radius $R_I = 70\, h^{-1}\,{\rm Mpc}$. To each ideal survey we apply a cut-off at $160\, h^{-1}\,{\rm Mpc}$, the survey limit.
\item The Maximum Likelihood Estimate (MLE).
\end{enumerate}
Our measurement represents an estimation of the bulk flow in the southern hemisphere, out to $50-70\mpch$.

\begin{table*}
\begin{center}
\caption{Bulk flow results for the MV and MLE estimators, assuming peculiar velocity uncertainties $\sigma_n = 0.324 H_0 D_z  \kms$ for each galaxy.  Columns are the bulk flow magnitudes $|{\bf U}|$, the vector components $(u_x,u_y,u_z)$,  and angular coordinates.  The top panel shows Galactic coordinates, with angles in Galactic longitude ($l$) and latitude ($b$), while the lower panel shows Equatorial coordinates, with angles in Right Ascension (RA) and Declination (Dec). The uncertainties quoted are noise, with the cosmic variance uncertainty in parentheses. The MV methods use an ideal Gaussian window function, with radius $R_I = 50h^{-1}\,{\rm Mpc}$ or $R_I = 70h^{-1}\,{\rm Mpc}$, and with a cut-off at $160h^{-1}\,{\rm Mpc}$ corresponding to the redshift cut-off of the survey.}
\label{table_BFresults_final_new}
\begin{tabular}{@{}ccccccccc|c}
\Xhline{2\arrayrulewidth}
Bulk flow &  $|{\bf U}|$ & $u_x$ & $u_y$ & $u_z$ & $l$ / RA & $b$ / Dec \\
estimator & (km s$^{-1}$) & (km s$^{-1}$) & (km s$^{-1}$) & (km s$^{-1}$) & ($^\circ$)& ($^\circ$)  \\
\hline
\textit{Galactic coordinates} \\
MV ($R_I = 50h^{-1}\,{\rm Mpc}$)  & $248 \pm 58(100)$ & $142 \pm 66(106)$ & $-127 \pm 72(114)$ & $159 \pm 59(103)$ & $318 \pm 20$ & $40 \pm 13$ \\
MV ($R_I = 70h^{-1}\,{\rm Mpc}$)  & $243 \pm 58(101)$ & $139 \pm 66(106)$ & $-125 \pm 72(114)$ & $154 \pm 59(102)$ & $318 \pm 20$ & $39 \pm 13$ \\
MLE & $295 \pm 48(138)$ & $43 \pm 56(130)$ & $72 \pm 52(165)$ & $283 \pm 47(129)$ & $59 \pm 36$ & $74 \pm 11$ \\
\hline
\textit{Equatorial coordinates} \\
MV ($R_I = 50h^{-1}\,{\rm Mpc}$)  & $248 \pm 58(100)$ & $-208 \pm 55(96)$ & $-99 \pm 63(101)$ & $-91 \pm 77(125)$ & $205 \pm 16$ & $-21 \pm 17$ \\
MV ($R_I = 70h^{-1}\,{\rm Mpc}$)  & $243 \pm 58(101)$ & $-203 \pm 55(95)$ & $-97 \pm 63(100)$ & $-90 \pm 78(124)$ & $205 \pm 16$ & $-22 \pm 18$ \\
MLE & $295 \pm 48(138)$ & $-212 \pm 46(114)$ & $-125 \pm 55(115)$ & $162 \pm 53(186)$ & $211 \pm 13$ & $33 \pm 10$ \\
\Xhline{2\arrayrulewidth}
 \end{tabular}
 \end{center} 
 \end{table*}

The results are presented in Table~\ref{table_BFresults_final_new}, in both Galactic Cartesian coordinates and Equatorial Cartesian coordinates. We include the Equatorial coordinates, since 6dFGSv covers only half the sky in the Equatorial $z$ direction (i.e. the southern hemisphere), and we would therefore expect increased variance in this direction; we wish to make any such effect clearly distinguishable.  We may expect a smaller variance in the $x$ and $y$ directions. The uncertainties quoted are the noise uncertainties, with cosmic variance in parentheses. The cosmic variance is predicted for a given $\Lambda$CDM power spectrum, as we discuss further in Section~\ref{section_results_cosmologicalmodels}. 
 
For the MV estimator with $R_I = 50h^{-1}\,{\rm Mpc}$, we find a bulk flow amplitude of $ |{\bf U}| = \magf \kms$ in the direction $(l,b) = (\glfcirc,\gbfcirc)$, and for $R_I = 70h^{-1}\,{\rm Mpc}$, we find a bulk flow amplitude of $|{\bf U}| = \mags \kms$ in the direction $(l,b) = (\glscirc,\gbscirc)$.

\begin{figure}
\includegraphics[width=9cm]{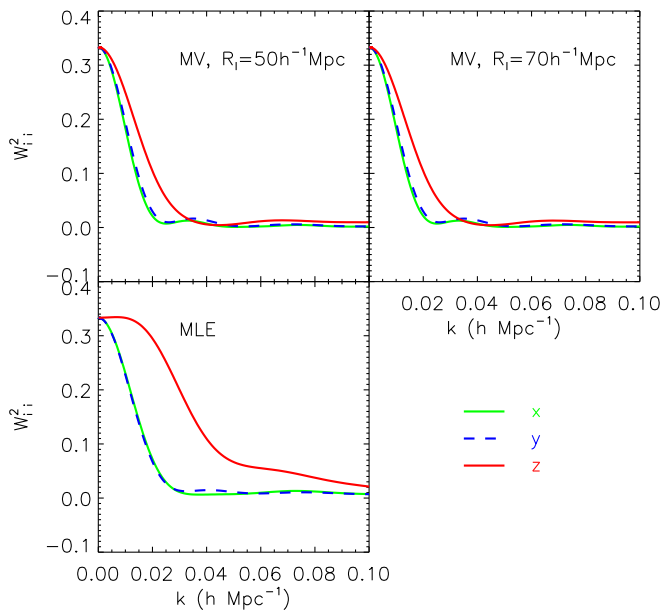}
\caption{The window functions $\mathcal{W}^2_{ii}$ (from Eq.~\ref{eqn_tensorwinfun}) of the bulk flow components for 6dFGSv, for each of our three estimators: the MV estimate with $R_I = 50$ or $R_I=70\,h^{-1}\,{\rm Mpc}$, and the MLE method. The Equatorial Cartesian $x,y,z$ components are the solid green, dashed blue, and solid red lines, respectively.}
\label{W2_grid}
\end{figure}

For the MLE, we find a bulk flow of ${ |{\bf U}| = 295\pm 48 \,{\rm km}\,{\rm s}^{-1}}$ in the direction $(l,b) = (59^\circ\pm36^\circ,74^\circ\pm11^\circ)$, which is not consistent with the direction of the MV results. The difference is largest in the Equatorial $z$ direction, and we can see why from looking at the window function $\mathcal{W}^2_{ii}$ of the different estimators, calculated from Eq.~\ref{eqn_tensorwinfun}, in Figure~\ref{W2_grid}. While the $x$ and $y$ window functions are similar for all the estimators, the $z$ window function is less compact for the MLE, giving more weight to smaller scales. 

We wish to clarify that the MV and MLE methods are different estimators of the bulk flow, and do not necessarily have to agree. They are based on different weightings over the volume, hence their different window functions, and so are quite free to give different results for both the amplitude and direction of the bulk flow. 
The MLE is much more sensitive to the window function of the survey than the MV, since the MV upweights a specified scale, while the scale of the MLE depends on the number of galaxies, their distribution, and their uncertainties. The 6dFGSv survey covers only half the range of scales in the Equatorial $z$ direction (i.e. the north-south direction) than the $x$ and $y$ directions (i.e. east-west), and so smaller scales contribute to the MLE bulk flow in the $z$ direction. This is why the MLE window function is less compact in the $z$-direction. 
There is significant variance in the small-scale 6dFGSv velocity field, as shown by S14, so a difference in window function can give quite a large difference in the bulk flow, which is what we observe. What is important is how we compare the results with a theoretical model. The MV method is more straightforward to compare with theory, since it gives the bulk flow for a specified window function which we can include in the theoretical model. 

\begin{figure*}
\includegraphics[width=13cm]{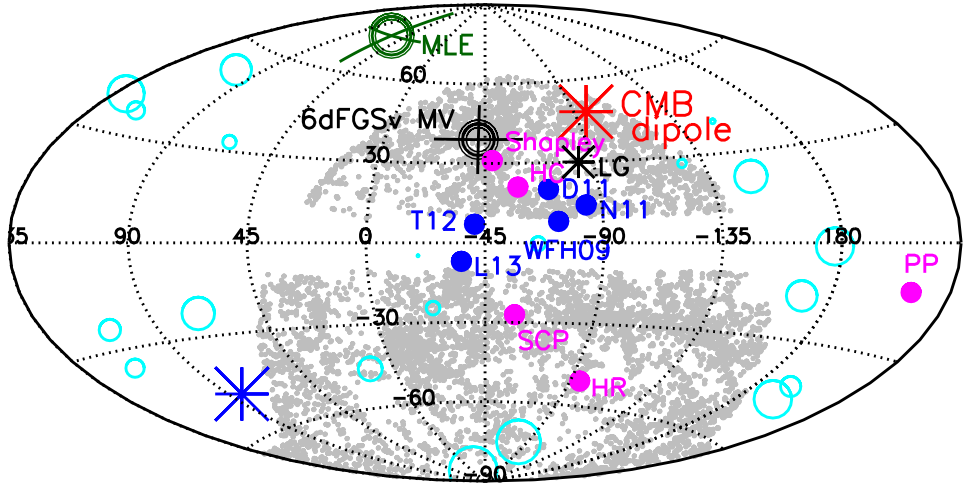}
\caption{ The 6dFGSv bulk flow result in this work, compared with other bulk flow measurements and nearby superclusters. The figure shows Galactic longitude ($l$) and latitude ($b$), in an Aitoff projection. Our MV result for $R_I=70h^{-1} \,{\rm Mpc}$ is shown as the black circle, while our MLE result is shown as the green circle (labelled). The diameter of the circles is proportional to the amplitude of the bulk flow, with inner and outer circles indicating the $1\sigma$ confidence interval of this amplitude.  The error bars show the $1\sigma$ angular uncertainty. The cyan circles show the distribution of bulk flows measured in our 20 GiggleZ 6dFGSv mock catalogues. Again, the size of the circles corresponds to the bulk flow amplitude, and since these are from simulations they have no measurement uncertainties. Our result for $R_I=50h^{-1} \,{\rm Mpc}$ is almost identical to the one for $70 \mpch$. The 6dFGSv galaxies are shown in grey for reference. We show the directions of several other results from the literature by the solid blue circles: WFH09 (W09), {\protect \cite{dai2011}} (D11), {\protect \cite{nusser2011}} (N11), {\protect\cite{turnbull2012}} (T12), {\protect \cite{lavaux2013}}  (L13) and {\protect \cite{planck2013_pv}} (P13). The four largest local superclusters are shown by the solid magenta circles: the Shapley Supercluster, Hydrus-Centaurus (HC), Horologium-Reticulum (HR) and Perseus-Pisces (PP). The South Celestial Pole (SCP) is also shown in magenta for reference. The CMB dipole is indicated by the red and blue stars (with red the direction of the dipole), and the direction of the Local Group motion \citep[from][]{kogut1993}  is shown by the black star. } 
\label{6dFGS_BFresults_galactic}
\end{figure*}

 We show the sky positions of our MV and MLE bulk flow measurements in Figure~\ref{6dFGS_BFresults_galactic}. The CMB dipole is shown for comparison, along with a number of recent bulk flow measurements from the literature.  We also show on this plot the position of the Shapley Supercluster ($l=312^\circ,b=31^\circ$). Our MV measurement is very close to the direction of Shapley, and consistent with it within the angular uncertainties. Unlike all-sky peculiar velocity surveys, 6dFGSv will be dominated by southern-sky structures, since the gradient of the velocity field towards these structures will be larger, and so it is not surprising that our measurement is close to Shapley.  Also, the 6dFGSv number density of galaxies peaks beyond $100\mpch$, incorporating part of Shapley, so this survey selection criteria itself will likely cause the Shapley region to dominate our bulk flow results.
 
In this figure, we also show the bulk flow results from our 20 GiggleZ-based mock catalogues, which we will discuss further in Section~\ref{section_results_mocks}. 

%----------------------------------------------------------
\subsection{6dFGSv bulk flow distribution in $\Lambda$CDM mocks}
\label{section_results_mocks}

\begin{figure}
\begin{center}
\includegraphics[width=9cm]{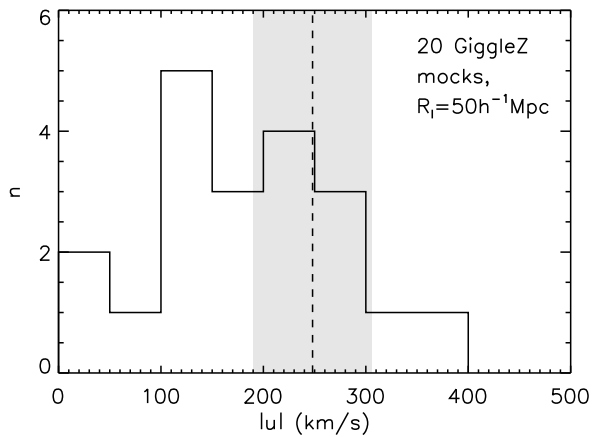}
\caption{Histogram of the bulk flow amplitudes $|\textbf{\textit{U}}|$ in our 20 GiggleZ-based 6dFGSv mocks, for the MV estimator with $R_I = 50h^{-1}\,{\rm Mpc}$. The vertical dashed line shows the corresponding amplitude for the data, and the grey shaded area indicates the $1\sigma$ noise uncertainty in the measurement.}
\label{BFhist_6dFGSv_Gz}
\end{center}
\end{figure}

We use our $N$-body simulation-based mock catalogues to determine the expected distribution of bulk flows for 6dFSGv in a $\Lambda$CDM universe. We calculate the bulk flow amplitude in each of the 20 mocks, using the MV method with $R_I = 50h^{-1}\,{\rm Mpc}$, and show their histogram in  Figure~\ref{BFhist_6dFGSv_Gz}.  We also show the corresponding bulk flow magnitude from the data, along with the $1\sigma$ noise uncertainty; this is above the average, but within the expected range of the mocks. Seven of the mocks, or $35\%$, lie above our result, while $65\%$ lie below.

The direction and amplitude of the bulk flow measured in each of these mocks is shown in Figure~\ref{6dFGS_BFresults_galactic}. This is a useful test to see whether the survey window function can bias the direction of the measured bulk flow. We see that the directions of the mocks appear fairly random and isotropic. There are more in the northern sky (13) than the southern sky (7), but this is not significant. It is also possible the mocks may share large-scale modes, since they all lie within the same Gpc volume, so it could be they are not completely independent. 

\subsection{Comparison with linear theory: 3D bulk flow}
\label{results_3d_theory}

Since the bulk flow amplitude is sensitive to the large-scale modes of the matter power spectrum, the measured bulk flow can be compared with the predicted value for a given cosmological model.  If the Universe is statistically homogeneous and isotropic, then the expected mean bulk flow at any location is zero. The root-mean-square (RMS) variance of the bulk flow amplitude, however, is cosmologically interesting, since it depends on the matter power spectrum, as well as the scale and window function in which it is measured.

\begin{figure*}
\includegraphics[width=18cm]{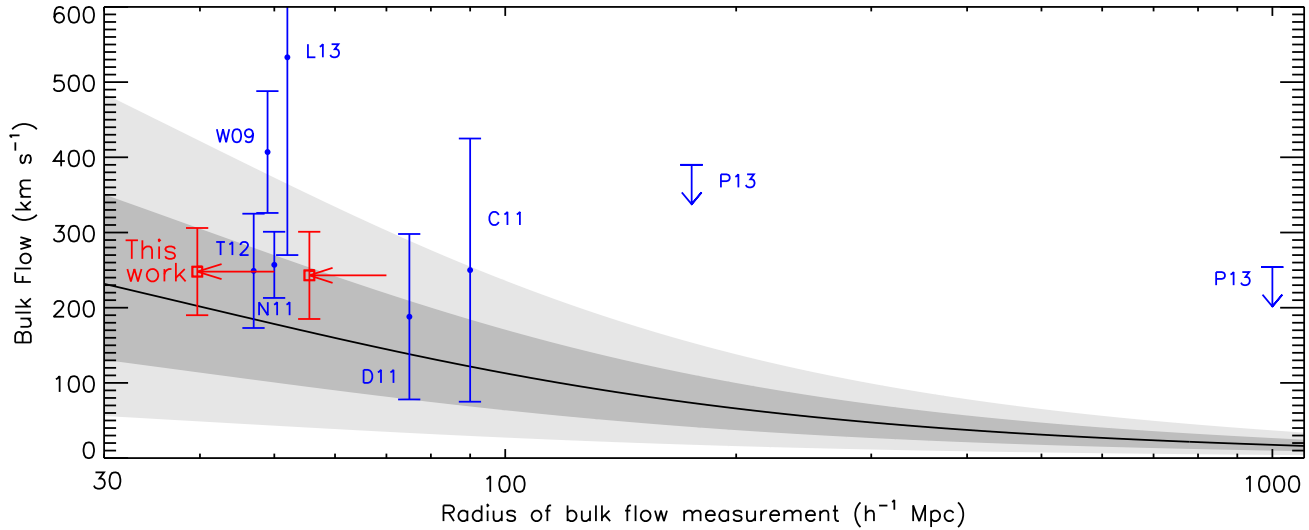}
\caption{The 6dFGSv bulk flow results in this work (red squares with error bars) for the MV method with radius $R_I=50$ and $70 \,h^{-1}\,{\rm Mpc}$, compared to a $\Lambda$CDM prediction. These results are plotted at `effective radii', corresponding to the radius of a full sphere with the same volume as the half-sky measurement, to show the variance we actually expect. The red arrows show how far we have shifted the points from the measured radii $R_I$ to the effective radii $R_{\rm eff}$.  The black solid curve is a linear-theory $\Lambda$CDM prediction for an all-sky Gaussian window function. The dark grey and light grey regions show the $68.3\%$ and $95.5\%$ confidence levels, assuming a Maxwellian distribution of velocities.  Other recent measurements are shown in in blue -- these are: {\protect \cite{lavaux2013}}  (L13), WFH09 (W09), {\protect\cite{turnbull2012}} (T12), {\protect \cite{colin2011}} (C11), {\protect \cite{planck2013_pv}} (P13), {\protect \cite{dai2011}} (D11), and {\protect \cite{nusser2011}} (N11). Several of these results - C11, D11, N11 and P13 - have top hat windows, and so we plot them at half their quoted radius, to be more comparable to the characteristic radius of the Gaussian window prediction. %\color{blue}\st{We also plot the M14 result at the effective radius of a full sphere, in the same way as our results.}\color{black}  
All error bars are $1\sigma$, while the two Planck arrows are the 95\% upper limits.} 
\label{pecdat}
\end{figure*}

We compare our 6dFGSv MV bulk flow amplitude, for both our ideal survey radii, to a $\Lambda$CDM linear-theory prediction in Figure~\ref{pecdat}. This prediction is the most likely bulk flow amplitude, $V_{\rm ML}(R) $, which depends on the RMS velocity dispersion, $\sigma_V$. 
The RMS velocity dispersion is given by
\begin{equation}
\label{eqn_rms_velocity}
\sigma_V^2(R) \equiv \langle V(R)^2\rangle = \frac{H_0^2f^2}{2\pi^2} \int_{k=0}^\infty {\rm d}k P(k) \widetilde{W}(k; R)^2,
\end{equation}
where $P(k)$ is the matter power spectrum, and $\widetilde{W}(k; R)$ is the Fourier Transform of the window function, $W(R)$, at effective radius $R$. In this plot, we use an all-sky Gaussian window function, $\widetilde{W}_G = \exp(-k^2R^2/2)$. (In the next section, we will use the exact window function of the survey to perform cosmological fits.)

The expected bulk flow velocity $V(R)$ can be predicted from $\sigma_V$, assuming the peculiar velocity field is Maxwellian, which it will be if the density field is Gaussian random. For a Maxwellian distribution, the probability distribution function of the bulk flow amplitude $V$ is \citep{bahcall1994,coles2002}
\begin{equation}
\label{eq_pdf_maxwellian}
p(V) {\rm d}V = \sqrt{\frac{2}{\pi}} \left( \frac{3}{\sigma_V^2} \right)^{3/2} V^2 \exp \left( - \frac{3V^2}{2\sigma_V^2} \right) {\rm d}V.
\end{equation}
For such a distribution the most likely (maximum likelihood) bulk flow amplitude is $V_{\rm ML}=\sqrt{2/3}\,\sigma_V$, while the expectation value is $\langle V \rangle = 2V_{\rm ML}/\sqrt{\pi} = \sqrt{8/3\pi}\,\sigma_V$. 

In Figure~\ref{pecdat} we plot $V_{\rm ML}$ along with the upper and lower $1\sigma$ and $2\sigma$ confidence levels as the dark and light grey shaded regions, found from integrating Equation~\ref{eq_pdf_maxwellian}. These confidence levels correspond to the variance ranges ${V_{\rm ML}}^{+0.419 \sigma_v}_{-0.356\sigma_v}$ ($1\sigma$) and ${V_{\rm ML}}^{+0.891 \sigma_v}_{-0.619\sigma_v}$ ($2\sigma$). To calculate $\sigma_V$, we use a $\Lambda$CDM matter power spectrum, generated using \textsc{camb} \citep{lewis2000} with nonlinear evolution calculated using \textsc{halofit} \citep{smith2003}, and with the parameters listed in Section~\ref{section_6dfgs_intro}.

Caution is needed in interpreting this plot, since the different surveys have different window functions, and so cannot be directly compared, either with each other or with the theoretical prediction for a perfect all-sky Gaussian.  A selection function tends to reduce the effective scale of a survey, which increases $\sigma_V$ and hence $V$ for that survey. However, simulations show that the PDFs of bulk flows depend primarily on $\sigma_V$, and not on the type of window function, and so assuming an effective radius for the window function used in the model (e.g. a Gaussian in our case) that reproduces the same $\sigma_V$ as the survey window function would allow a comparison at that scale \citep{li2012}. We have not done this in this plot, and note that there is some uncertainty on the effective scale of the different surveys.

Since 6dFGSv only covers half the sky, we would expect our measurements at given radius $R_I$ to have more cosmic variance than predicted by the full-sky model at this radius. Conversely, we could consider 6dFGSv to be at a smaller effective radius.
We therefore plot our 6dFGSv MV results at `effective radii' $R_{\rm eff}$ accounting for the fact that 6dFGSv covers only half the sky.
For each of the $R_I = 50$ and $R_I = 70 \, h^{-1}\,{\rm Mpc}$ results, we calculate the radius of a full sphere with the same volume as the half-sky measurement, i.e.
\begin{equation}
R_{\rm eff} = (R_I^3/2)^{1/3}.
\end{equation}
This gives effective radii of $R_{\rm eff} = (39.7,55.6)\mpch$ for the $R_I = (50, 70)\mpch$ measurements. We plot arrows showing how we have shifted the measurements from $R_I$ to $R_{\rm eff}$.
However, since 6dFGSv is not a perfectly sampled hemisphere, we might expect the effective radii to be even smaller than the $R_{\rm eff}$ we calculate. 

From Figure~\ref{pecdat} we see that once shifted to the effective radii, both the 6dFGSv $R_I = 50 \, h^{-1}\,{\rm Mpc}$ and  $R_I = 70 \, h^{-1}\,{\rm Mpc}$ bulk flow results appear to be consistent within $68.3\%$ confidence with the theoretical prediction. 

The uncertainty on the effective radii of previous surveys may mean that those that showed higher than predicted bulk flows could have been compared to theory at too large a radius, without accounting for how the window function reduces the effective volume of the survey. It would be illuminating to recalculate the effective radii of these surveys to investigate this; we leave this for future work.

%----------------------------------------------------------
\subsection{Comparison with linear theory: 1D bulk flow}
\label{section_results_cosmologicalmodels}

Unlike the 3D bulk flow amplitude, the 1D bulk flow components $u_i$ are Gaussian-distributed, making them more useful for a robust test of $\Lambda$CDM.
The 1D RMS velocity variance is given for a particular survey by the covariance matrix of the bulk flow moments, $R_{ij}$, which we defined in section~\ref{section_BF_uncertainties}. 
It is the sum of a noise component and a cosmic variance component, and it depends on the survey geometry, the measurement noise, and the matter power spectrum. It is very similar to the RMS velocity variance $\sigma_V$ in Equation~\ref{eqn_rms_velocity}, except for the addition of the noise component, and the cosmic variance component $R_{ij}^{(v)}$ contains the tensor window function $\mathcal{W}_{ij}(k)$.  (We previously defined $\sigma_*$ as the 1D velocity variance; this is in principle the average variance over all scales, which we assumed to be equal to $\sim 250 \kms$. Here, however, we are looking at the variance as a function of scale.) 

The deviation from zero of the observed bulk flow components $u_i$ can be directly compared with the predicted dispersion, by calculating the $\chi^2$ for the three moments, 
\begin{equation}
\label{eqn_bf_chi2}
\chi^2 = \sum_{i,j} u_i R_{ij}^{-1} u_j,
\end{equation}
where $i$ and $j$ both go from 1 to 3 to specify the bulk flow components, $u_i$ and $u_j$ are the measured bulk flow components, and $R_{ij}$ is the covariance matrix of the moments for a specified set of cosmological parameters. 

 $R_{ij}$ is dominated by the cosmic variance term, typically of order $\sim100\,{\rm km}\,{\rm s}^{-1}$, while the noise term is typically $\sim40\,{\rm km}\,{\rm s}^{-1}$. Since the bulk flow depends on large-scale density fluctuations, $R_{ij}$ will be most sensitive to the amplitude and shape of the power spectrum. The power spectrum amplitude is parameterised by  the RMS density fluctuations in spheres of $8 h^{-1}\,{\rm Mpc}$ radius, $\sigma_8$, while the shape is parameterised by the shape parameter, $\Gamma$, which on large scales can be approximated by $\Gamma=\Omega_{\rm m}h$. The dependence on $\Omega_{\rm m}$ also comes into the $f(\Omega_{\rm m},z)^2$ factor. We therefore follow WFH09 in using the bulk flow to constrain a combination of $\Omega_{\rm m}$ and $\sigma_8$ -- in our case, we fix $h$ to the best-fit value from \textit{Planck}, $h=0.67$. 
 
 In order to fit $\Omega_m$ and $\sigma_8$ we use the likelihood, following WFH09, which is given by 
 % Since the covariance matrix $R_{ij}$ depends on the fitted parameters, the likelihood depends on the determinant of the covariance, and so to fit parameters we need to use the likelihood rather than the $\chi^2$. The likelihood is given by
\begin{equation}
\mathcal{L}(\Theta) \propto \frac{1}{\sqrt{|R|}} \exp \left( \sum_{i,j} -\frac{1}{2}u_iR_{ij}^{-1}u_j \right),
\end{equation}
 where $\Theta$ is the vector of parameters. In our case, $\Theta=(\Omega_m,\sigma_8)$, and we fix all other parameters to their \textit{Planck} values. %This equation is similar to Equation~\ref{eqn_bf_chi2} except that it is normalised by the factor $\ln|R|$, which down-weights models with high cosmic variance. This allows us to obtain closed constraints on $\Omega_m$ and $\sigma_8$.

\begin{figure}
\includegraphics[width=9cm]{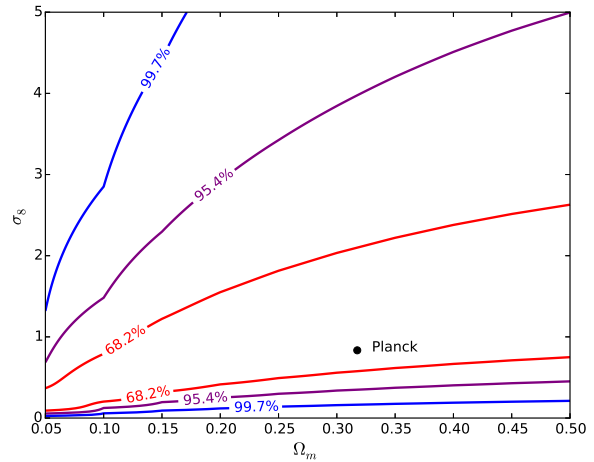}
\caption{The likelihood-based confidence levels on $\Omega_{\rm m}$ and $\sigma_8$, obtained from the 6dFGSv MV bulk flow measurement with $R_I = 70 \mpch$. The black point indicates the best fit values found by \textit{Planck}, used as the fiducial values in this work. The result for $R_I = 50 \mpch$ is very similar to this plot. }
\label{OM_sig8_contour_6dFGS_MV50DW}
\end{figure}

We show our constraints on $\sigma_8$ and $\Omega_{\rm m}$ for our MV results in Figure~\ref{OM_sig8_contour_6dFGS_MV50DW}. There is a degeneracy between $\sigma_8$ and $\Omega_{\rm m}$, since a lower $\sigma_8$ requires a lower $\Omega_{\rm m}$ to produce the same bulk flow; or, for fixed $\sigma_8$, lower values of $\Omega_{\rm m}$ lead to a larger bulk flow. This is because if $\sigma_8$ is fixed, then a lower $\Omega_{\rm m}$ requires a larger power spectrum amplitude on large scales to allow this normalisation.  However, since a lower $\Omega_{\rm m}$ also decreases the growth rate $f(\Omega_{\rm m},z)$, these two effects partially cancel, and so the bulk flow does not have much constraining power on $\Omega_{\rm m}$. The $\textit Planck$ value is shown as the black point, and is within the $1\sigma$ range of our measurement.

%\color{blue}
%\st{Our measurement is much more sensitive to $\sigma_8$, but it can only give a lower bound, since for any $\sigma_8$ there is no lower bound on the expected bulk flow. The bulk flow measurement is larger than the $\Lambda$CDM prediction, and clearly favours a high $\sigma_8$, though it is consistent with the \textit{Planck} value within $2\sigma$. For the \textit{Planck} parameters, we find $\chi^2=4.23$ for our MV $R_I=50$ measurement, and $\chi^2=5.21$ for our MV $R_I=70$ measurement, for 3 degrees of freedom.}
%\color{black}

%{\color{blue} \st{WFH09 find that their bulk flow measurement requires a very high $\sigma_8$ of $\sim1.7$, with lower 95 and 99 per cent limits of $1.11$ and $0.88$, respectively. To compare to their result, following their method}} 
%Like WFH09 we calculate the likelihood of $\sigma_8$, fixing all other parameters to their \textit{Planck} values, using
%\begin{equation}
%\mathcal{L}(\sigma_8) \propto \frac{1}{\sqrt{|R|}} \exp \left( \sum_{i,j} -\frac{1}{2}u_iR_{ij}^{-1}u_j \right).
%\end{equation}
%This is similar to Equation~\ref{eqn_bf_chi2} except that it is normalised by the factor $\ln|R|$, which down-weights models with high cosmic variance.

\begin{figure}
\includegraphics[width=9cm]{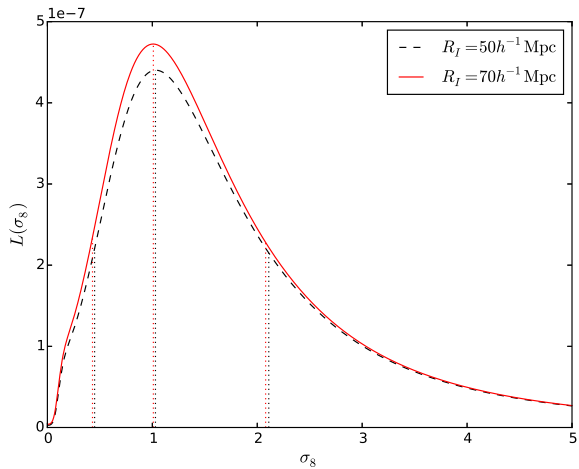}
\caption{ Likelihood of the value of $\sigma_8$ from our bulk flow measurement, after marginalising over $\Omega_m$. The black dashed curve is for the MV $R_I=50\,h^{-1}\,{\rm Mpc}$ measurement, and the red curve is for $R_I=70\,h^{-1}\,{\rm Mpc}$. The dotted lines indicate the maximum of the likelihood and $68.27\%$ confidence levels.}
\label{likelihood_s8}
\end{figure}

Marginalising over $\Omega_m$, we obtain the likelihood for $\sigma_8$. The results are shown in Figure~\ref{likelihood_s8}. Our results favour a high value of $\sigma_8$, but we do not find a significant disagreement with $\Lambda$CDM. For the MV $R_I=50\mpch$ measurement, we find %$\sigma_8=1.08_{-0.51}^{+1.19}$ ($68.27\%$ C.L.)
$\sigma_8=1.03^{+1.08}_{-0.58}$ ($68.27\%$ C.L.), and for $R_I=70\mpch$, we find %$\sigma_8=1.05_{-0.50}^{+1.18}$
$\sigma_8=1.01^{+1.07}_{-0.58}$. Both of these are consistent with the \textit{Planck} value of $0.83 \pm 0.03$ \citep{planck2013_xvi}.

The diagonal elements of $R_{ij}$ provide the expected 1D RMS bulk flow variance $\sigma_{V,i}^2$ in each of the three directions $i$, for a given survey window function and model power spectrum.  Following WFH09, we list $\sigma_{V,i} = \sqrt{R_{ii}}$ in Table~\ref{table_pred_sigmav_6dfgsv} for 6dFGSv, for the MLE and MV estimators.  

 \begin{table*}
\begin{center}
\caption{Comparison of the expected 1D RMS velocity $\sigma_{V,i}$, and 3D RMS velocity $\sigma_{V}$, for 6dFGSv, our GiggleZ-based 6dFGSv mocks, and for theory.  We use Equatorial Cartesian coordinates, and assume a $\Lambda$CDM cosmology with parameters listed in Section~\ref{section_6dfgs_intro}.} 
\label{table_pred_sigmav_6dfgsv}
\begin{minipage}{\textwidth}
\begin{tabular}{cccccc}
\hline
\multicolumn{2}{c}{Source \&}  & $\sigma_{V,x}$ & $\sigma_{V,y}$ & $\sigma_{V,z}$ & $\sigma_V$ \\
\multicolumn{2}{c}{$R_I$  $(h^{-1}\,{\rm Mpc})$} & (${\rm km}\,{\rm s}^{-1}$) & (${\rm km}\,{\rm s}^{-1}$) & (${\rm km}\,{\rm s}^{-1}$) & (${\rm km}\,{\rm s}^{-1}$) \\
\hline
6dFGSv:\footnote{$\sigma_{V,i}=\sqrt{R_{ii}}$, from Eq.~\ref{eqn_Rij_full}, and  $\sigma_V^2 = J\, R_{ij} J^{\rm T}$, where J is the Jacobian. Includes both noise and cosmic variance.} 
& MLE ($R_I \sim 70$) & 122 & 122 & 193 & 139 \\
& MV, $R_I = 50$ & 95 & 100 &122 & 101  \\
& MV, $R_I = 70$ & 95 & 100 & 123 &102  \\
\hline
Mocks:\footnote{All calculated from root mean square of bulk flow components of 20 mocks. Includes both noise and cosmic variance.} &MV, $R_I = 50$ & 129 & 114 & 120 & 210 \\
\hline
Theory:\footnote{Calculated from Eq.~\ref{eqn_rms_velocity}. Includes noise only and assumes a full-sky window function.} &$\widetilde{W}_G$, $R_I = 50$ &- & -& -& 218   \\
&$\widetilde{W}_G$, $R_I = 70$ &- &- &- & 177  \\
\hline
 \end{tabular}
\end{minipage}
 \end{center} 
 \end{table*}

We also show in this table the 1D and 3D RMS velocities calculated from our 20 GiggleZ-based mocks, using the MV estimator with $R_I = 50 \, h^{-1}\,{\rm Mpc}$. We would expect these to closely agree with the 6dFGSv results for $R_I = 50\mpch$, since the mocks reproduce the window function of the data, and this is roughly true.  The last two rows of Table~\ref{table_pred_sigmav_6dfgsv} show   the analytic prediction for an all-sky Gaussian window function with radius $50$ or $70\,h^{-1}\,{\rm Mpc}$, by evaluating Equation~\ref{eqn_rms_velocity} at these values of $R$. 

\subsection{Bulk flow in redshift shells}

\begin{figure}
\includegraphics[width=9cm]{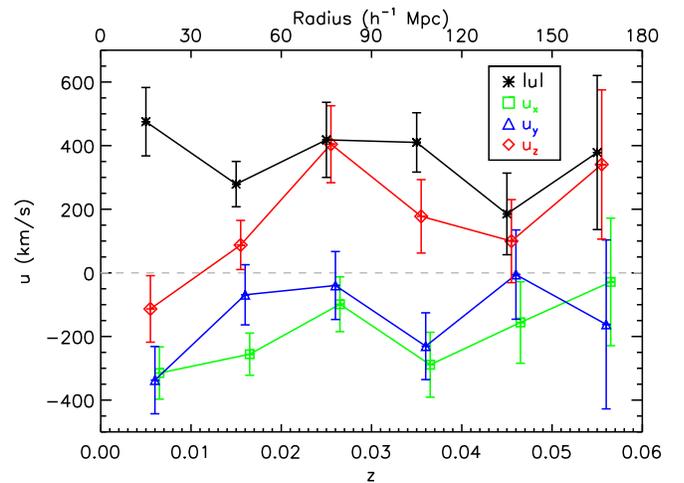}
\caption{The MLE bulk flow for 6dFGSv in redshift shells of width $\Delta z = 0.01$. The coloured lines show the Equatorial $(x,y,z)$ components (labelled) and are shifted to the right for clarity. The black line shows the bulk flow magnitude.  The number of galaxies in each redshift shell is $\{75,813,1371,2563,2802,1261\}$. The error bars indicate the noise uncertainty in each bin.} 
\label{6dFGSv_zbinMLE_directv_DW}
\end{figure}

It is interesting to look at how the bulk flow varies as a function of redshift. In Figure~\ref{6dFGSv_zbinMLE_directv_DW} we plot the MLE bulk flow, split into redshift bins of $\Delta z = 0.01$. In each redshift bin, we re-calculate the MLE weights for only galaxies in that bin. The results are noisy, but the amplitude of the bulk flow seems to be fairly constant up to the maximum redshift of $z=0.054$. This is what would be expected if the source of the bulk flow is an overdensity more distant than the scales measured.

\subsection{Zero-point uncertainty}

So far our analysis has assumed a fixed zero-point for the Fundamental Plane. \cite{springob2014} fix the zero-point by assuming zero average radial peculiar velocity in a `great circle' around the equator, consisting of 3828 galaxies with $-20^\circ \le {\rm Dec} \le 0^\circ$. (In practice they assume zero average logarithmic distance ratio $\eta$).  However, this zero-point estimation has both statistical uncertainty and cosmic variance. We investigate the effect of these on our bulk flow measurement here.

The statistical uncertainty on the zero-point was calculated by \cite{springob2014} to be 0.003 dex.  We test the effect of this uncertainty on our bulk flow measurement, by repeating the analysis but first shifting all the $\eta$ values by $+ 0.003$ dex or $-0.003$ dex. This changes the derived peculiar velocities $v$, the estimated velocity uncertainties $\sigma_n$ (see Section~\ref{deriving_vpec}), and the resulting bulk flow. We list the new $\sigma_n$ and MV bulk flow values in Table~\ref{table_zeropoint_sv}.

\begin{table*}
\begin{center}
\caption{The effect of statistical uncertainty in the Fundamental Plane zero-point on the 6dFGS bulk flow measurements. The best-fit velocity uncertainty $\sigma_n$, and MV bulk flow magnitude $|{\bf U}|$ and Galactic latitude and longitude $(l, b)$, are listed for the 6dFGSv measurement after adding or subtracting the statistical uncertainty on the zero-point, $0.003$ dex, from all the logarithmic distance ratios $\eta$.}
\label{table_zeropoint_sv}
\begin{tabular}{cccccc}
Quantity measured & \multicolumn{3}{c}{Amount by which we shift $\eta$ values of all galaxies} \\
& $-0.003$ dex &  0 dex (result) & $+0.003$ dex \\
\hline
$\sigma_n$  & $0.326H_0D_z$ & $0.324 H_0D_z$ & $0.322H_0D_z$ \\
\hline
MV BF, $R_I=50\mpch$ \\
$|{\bf U}|$ $(\kms)$ & $238\pm55(97)$ & $248\pm58(100)$ & $266\pm62(105)$ \\
$(l,b)$ $(^\circ)$ & $(324\pm25,51\pm14)$ & $(318\pm20,40\pm13)$ & $315\pm16,30\pm12$ \\
\hline
MV BF, $R_I=70\mpch$ \\
$|{\bf U}|$ $(\kms)$ & $231\pm55(97)$ & $243\pm58(101)$ & $263\pm62(105)$  \\
$(l,b)$ $(^\circ)$ & $(324\pm26,51\pm14)$ & $(318\pm20,39\pm13)$ & $(314\pm16,30\pm12)$\\
\hline
\end{tabular}
\end{center}
\end{table*}

\begin{table*}
\begin{center}
\caption{Uncertainty in the 6dFGS MV and MLE bulk flow measurements due to cosmic variance in the Fundamental Plane zero-point. We list the RMS variance found from our mocks in Equatorial Cartesian coordinates $(u_x,u_y,u_z)$, Galactic coordinates $(l,b)$, Equatorial coordinates (RA, Dec) and bulk flow magnitude $|{\bf U}|$.}
\label{table_zeropoint_cv}
\begin{tabular}{ccccccccc}
& $\delta u_x$ & $\delta u_y$ & $\delta u_z$ & $\delta l$ & $\delta b$ & $\delta$RA & $\delta$Dec & $\delta |{\bf U}|$ \\
& $(\kms)$ & $(\kms)$ & $(\kms)$ & $(^\circ)$ & $(^\circ)$ & $(^\circ)$ & $(^\circ)$ & $(\kms)$ \\ 
\hline
MV BF, $R_I=50\mpch$ & 3 & 8 & 128 & 28 & 21 & 2 & 25 & 51 \\
MV BF, $R_I=70\mpch$ & 3 & 8 & 128 & 30 & 22 & 2 & 25 & 52 \\
\end{tabular}
\end{center}
\end{table*}

The changes in the MV bulk flow values we find, after adding or subtracting $0.003$ dex, are all smaller than the noise uncertainties on our bulk flow measurement. Hence the statistical uncertainty on the zero-point does not significantly affect our measurement.

There is also cosmic variance in the zero-point, i.e. in the net velocity of galaxies within the `great circle', $-20^\circ \le {\rm Dec} \le 0^\circ$. We can estimate this using our $\Lambda$CDM mock catalogues, as follows:
\begin{enumerate}
\item In each mock catalogue, calculate the mean radial component of peculiar velocity of galaxies within $-20^\circ \le {\rm Dec} \le 0^\circ$.
\item Then subtract this mean radial velocity from all of the galaxies in the mock.
\item Calculate the MLE bulk flow of the mock before and after doing this. 
\item Calculate the vector difference between these.
\end{enumerate}
We find the RMS variance in the mean radial peculiar velocity in the great circle, over all the mocks, is $82 \kms$, and the RMS variance in logarithmic distance ratio $\eta$ is $0.004$ dex.  The RMS variance in the vector shift in bulk flow, over all the mocks, is $(\delta u_x, \delta u_y, \delta u_z)$ = $(3,8,128)\kms$ in Equatorial cartesian coordinates, shown in Table~\ref{table_zeropoint_cv}. In other words, the zero-point calibration induces a RMS variance primarily in the north-south direction. 

We note that if the net radial peculiar velocity in the `great circle' is negative, i.e. towards the observer, then calibrating it to zero shifts the measured bulk flow to more negative Dec. If the net velocity is positive, then the shift is towards more positive Dec. Either way, there is negligible shift in RA, since 6dFGSv is fairly symmetrical in Equatorial $x$ and $y$. 

The cosmic variance this adds to our 6dFGSv bulk flow measurement depends on the direction and amplitude of the measurement. We show the resulting cosmic variance on the amplitude and direction of our MV bulk flow measurements at $R_I = 50$ and $70\mpch$ in Table~\ref{table_zeropoint_cv}. The uncertainty on the bulk flow amplitude due to cosmic variance in the zero-point is $\sim50\kms$, slightly smaller than the statistical uncertainty of $58\kms$.

\subsection{Comparison with other results}

We compare our bulk flow result to other recent measurements in the literature, in Table~\ref{table_BF_summary}. Our result is one of the most precise to date, thanks to the large number of galaxies in 6dFGSv. Our MV result of $\magf \kms$ at $R_I = 50 \, h^{-1}\, {\rm Mpc}$ is a significantly lower amplitude than that of WFH09 at the same scale, despite the fact that the 6dFGSv survey volume is smaller than the COMPOSITE sample that they use, and so might be expected to have more cosmic variance. The level of disagreement between our result and WFH09, not accounting for this volume difference, is $1.56\sigma$. Our measurement also does not appear to support the high-redshift $600-1000\,{\rm km}\,{\rm s}^{-1}$ measurement of \cite{kashlinsky2008}, although since their scale is much larger we cannot directly rule it out.

Our result is consistent with a growing number of recent measurements that find a bulk flow amplitude consistent with $\Lambda$CDM, including \cite{colin2011}, \cite{dai2011}, \cite{nusser2011}, \cite{turnbull2012},  \cite{feindt2013} and \cite{carrick2014}. 

As we see in Table~\ref{table_BF_summary} and Figure~\ref{6dFGS_BFresults_galactic}, the direction of our bulk flow is much closer to Shapley than other bulk flow measurements. This is reasonable, since 6dFGSv covers only the southern hemisphere, and so the bulk flow we measure is likely to be dominated by large southern structures such as Shapley.

We note again that the different surveys quoted in this
table all have different window functions, so even those at
the same effective distance may not be directly comparable.
In particular a region of a ``quiet'' Hubble flow has been
identified in the northern sky \citep{courteau1993} which
is in contrast to the southern sky that has large motions
arising from the Great Attractor and the Shapley Supercluster
\citep[see e.g.][]{hudson1999,feindt2013}.
If the \cite{turnbull2012} SNIa dataset is
sub-divided into north and south samples, the measured bulk flow amplitudes are $110\pm90$ and 
$320\pm120 \kms$ respectively. As datasets improve such biases will need to be
fully addressed.

\begin{table*}
\begin{center}
\caption{Summary of some recent bulk flow results in the literature, compared to the result in this work. For each measurement, we list the distance indicator used (DI), the number of peculiar velocities in the sample $N$, the radius of the measurement $R$,  the measured bulk flow magnitude $|{\bf U}|$, and the direction of the bulk flow in Galactic longitude $l$ and latitude $b$. A dash for the DI means a combination of datasets were used -- these results all used the COMPOSITE sample. For measurements of the kSZ effect, $N$ shows the number of clusters used in combination with the CMB (with the exception of Lavaux (2013), who use galaxies instead of clusters). A number of these results use the same, or overlapping, datasets, but apply different analyses, and the window functions differ for each survey.}
\label{table_BF_summary}
\begin{minipage}{\textwidth}
\begin{tabular}{@{}ccccccccc|c}
\hline
& DI & N & R & $|{\bf U}|$ & $l$ & $b$ \\
 &&&  $(h^{-1}\, {\rm Mpc})$   & (${\rm km}\,{\rm s}^{-1}$) & ($^\circ$)& ($^\circ$) \\
\hline
6dFGSv (this work) & FP & 8885  & 50& $\magf$ & $\glf$ & $\gbf$ \\
 & FP & 8885 & 70& $\mags$ & $\gls$ & $\gbs$  \\
\hline
Dressler et al. (1987a) & FP & 423 & $\lesssim 60$& $599\pm104$  & $312\pm11$ & $6 \pm 10$ \\
Watkins et al. (2009) & Mix & 4481 & 50& $407\pm81$ & $287\pm9$ & $8\pm6$  \\
Feldman et al. (2010) & Mix & 4536 & 50& $416\pm78$ & $282\pm11$ & $6\pm6$ \\
\cite{macaulay2012} & Mix & 4537& 33  & $380^{+99}_{-132}$ & $295\pm18$ & $14\pm18$ \\
Ma \& Scott (2013) & Mix & 3304 & 50 & $340 \pm 40 $ & $280 \pm 8$ & $5.1 \pm 6$ \\
Nusser \& Davis (2011) & TF & 2859 & 40& $333\pm38$ & $276\pm3$ & $14\pm3$\\
Ma \& Pan (2014) & TF & 2915 & 58 & $290 \pm 30$ & $281\pm7$ & $8^{+6}_{-5}$ \\
Colin et al. (2011) & SNe & 142& 160& $260\pm150$ & $298^{+62}_{-48}$ & $8^{+34}_{-52}$\\
Dai et al. (2011) & SNe & 132& 150& $188^{+199}_{-103}$ & $290^{+39}_{-31}$ & $20{\pm 32}$ \\
Turnbull et al. (2012) & SNe & 254& 50& $249\pm76$ & $319\pm18$ & $7\pm14$\\
Feindt et al. (2013)\footnote{The result for their lowest redshift shell} & SNe & 128 & 74& $243\pm88$ & $298\pm25$ & $15\pm20$  \\
\cite{weyant2011} & SNe & 30& 112& $446\pm101$ & $273\pm11$ & $46\pm8$ \\ 
Kashlinsky et al. (2008) & kSZ & 782 & $\sim300-800$& $\sim600-1000$ & $283\pm14$ & $12\pm14$ \\
Planck Collaboration (2013) & kSZ & 1405 & 350& $<390$ (95\% CL) \\
&&& 2000  & $<254$ (95\% CL) & $\sim120$ & $\sim34$\\
Lavaux (2013) & kSZ & 5290 & 50 & $533 \pm 263$ & $324 \pm 27$ & $-7 \pm 17$ \\
&&& 200 & $284\pm187$ & $26\pm35$ & $-17 \pm 19$ \\
&&& $\sim500$ & $<470$ (95\% CL) \\
\hline
\end{tabular}
\end{minipage}
\end{center} 
\end{table*}

As a final point, recently \cite{johnson2014} measured the velocity power spectrum of the 6dFGSv dataset as a function of scale, and found it to be $1\sigma$ larger than the prediction given by a \planck cosmology on the largest scale they measured ($k=[0.005,0.02]$). However, they find it to be consistent at the $2\sigma$ level. This is consistent with our fit to $\Omega_{\rm m}$ and $\sigma_8$ in Figure~\ref{OM_sig8_contour_6dFGS_MV50DW}, which is also consistent with \planck at the $1\sigma$ level. As \cite{johnson2014} mention, this is not a significant disagreement with $\Lambda$CDM.

\subsection{Implications for Cosmography}

An important aim for bulk flow measurements has been to understand the motion of the Local Group (LG) with respect to the CMB, of $627\pm22 \,{\rm km}\,{\rm s}^{-1}$ towards $l=276\pm3^\circ,b=30\pm2^\circ$ \citep{kogut1993}.  From gravitational instability theory, this is expected to be caused by nearby structures, and to converge to the CMB dipole beyond them. 

As we showed in Figure~\ref{6dFGS_BFresults_galactic}, the direction of our bulk flow is consistent with the direction of the Shapley Supercluster.  We also saw in Figure~\ref{6dFGSv_zbinMLE_directv_DW} that the amplitude of the bulk flow remains fairly constant with distance, indicating that it is sourced by a distant rather than a nearby overdensity.  This therefore seems to indicate that Shapley may be the dominant source of the bulk flow motion we detect. Shapley is at a distance of $152\,h^{-1}\,{\rm Mpc}$, and is the largest supercluster in the local Universe out to $200\,h^{-1}\,{\rm Mpc}$ \citep{lavaux2011}.  Our result is consistent with many other bulk flow measurements that find directions close to Shapley \citep[e.g.][]{feindt2013} and a source distance greater than $\sim50-80\,h^{-1}\,{\rm Mpc}$ as the origin of the flow \citep[e.g.][]{hudson1994IV,kocevski2004,pike2005,watkins2009}.

\cite{lavaux2011} calculate, using linear theory applied to 6dFGS redshift data, that Shapley should be responsible for ${\sim15}$ per cent of the total velocity of the LG with respect to the CMB, or $90\pm10\,{\rm km}\,{\rm s}^{-1}$, while the Horologium-Reticulum supercluster generates ${\sim60\,{\rm km}\,{\rm s}^{-1}}$.  However, it appears that our sample is dominated mostly by Shapley. This makes it possible that its mass could be even larger than inferred from redshift data alone, which would agree with the finding of \cite{feindt2013}, who find that the bulk flow does not appear to reverse beyond Shapley, suggesting there could be more mass beyond it sourcing the bulk flow. They calculate that their bulk flow would be caused either if the mass of Shapley were twice as large as current estimates \citep[from][]{munoz2008,sheth2011}, or if there were a more distant mass behind Shapley.

As we have previously noted however, 6dFGSv partially samples the Shapley region,  with no sampling at all of northern-sky structures, so this could be partially responsible for Shapley dominating our results. Additionally, \cite{springob2014} show that the 6dFGSv sample shows not only an excess of positive velocities towards Shapley, but also an excess of negative velocities on the other side of the sky towards the Cetus Supercluster, compared to model predictions, indicating other structures are also contributing to the velocity dipole of the sample. As we found in Section 6.6, the cosmic variance in the zeropoint of the Fundamental Plane also gives additional angular uncertainty to our measurement in the north-south direction. More analysis would therefore be needed to confirm whether the bulk flow is truly closer to Shapley than any other structure. 

%***********************************************************************************************************************************************
\section{Conclusion}
\label{section_6dfgs_conclusion}

The question of whether a large bulk flow exists in the local Universe remains of much interest. A large part of the disagreement between previous measurements is likely due to the noisy, sparse peculiar velocity samples to date, as well as possible unknown systematics such as differently-calibrated datasets and Malmquist (or selection) biases. In this paper we aimed to make an improved measurement using a large new peculiar velocity dataset, the 6 degree Field Galaxy Survey peculiar velocity sample (6dFGSv). This sample is homogeneously selected, so avoids any bias from combining datasets, and the uncertainties and Malmquist biases have been carefully studied and accounted for (M12, S14). 

We have presented a new bulk flow analysis using this dataset. Using the `Minimum Variance' bulk flow estimator, we find a bulk flow of magnitude $|{\bf U}| = \magf\kms$ in the direction $(l,b) = (\glfcirc,\gbfcirc)$ at a distance of $50 \, h^{-1} \,{\rm Mpc}$, and $|{\bf U}| = \mags \kms$ in the direction $(l,b) = (\glscirc,\gbscirc)$ at a distance of $70 \, h^{-1} \,{\rm Mpc}$. This is somewhat higher than the $\Lambda$CDM prediction on these scales, implying a high value of $\sigma_8$, but consistent with \textit{Planck} results within $2\sigma$. After marginalising over $\Omega_{\rm m}$, we find from our bulk flow measurement at $R_I=70\mpch$ a value of $\sigma_8=1.01_{-0.58}^{+1.07}$, consistent with the \textit{Planck} value of 0.83 within $68.27\%$ confidence. 

Our result is in agreement with a number of recent measurements that also find a bulk flow consistent with $\Lambda$CDM, including \cite{turnbull2012}, \cite{feindt2013} and
\cite{hong2014}. Our result is also supported by the higher-redshift measurement of \cite{planck2013_pv}, who used \textit{Planck} CMB data combined with a large X-ray cluster catalogue, and found no evidence for a bulk flow from $350\mpch$ to $2\,h^{-1}\,{\rm Gpc}$ scales.

A challenge for the 6dFGSv analysis here (and for any peculiar velocity analysis made using linear velocities instead of log distances) is accounting for the lognormal uncertainties on the peculiar velocities. When combined with the fact that 6dFGSv only covers half the sky, these can result in a spurious polar bulk flow component if not properly accounted for. We have shown that it is important to propagate uncertainty from the Gaussian observable (in our case, the logarithmic distance ratio $\eta=\log_{10}D_z/D_r$) to the non-Gaussian velocity in a way that is independent of the $\eta \to v$ conversion itself, so that the velocity uncertainties, and hence bulk flow weights, do not correlate with the velocities. A further effect may come from the fact that the distribution of measured velocities themselves will be affected by the lognormal uncertainties. A possible solution to this problem was recently suggested by \cite{watkinsfeldman2015}. We leave investigation of this for 6dFGSv to future work.

Our measured bulk flow is very close to the direction of the Shapley Supercluster, consistent with many other measurements, and its amplitude appears to be fairly constant out to the distance of Shapley. This suggests that a large part of the bulk flow we measure is likely to be sourced by Shapley, which is reasonable since 6dFGSv is a southern-sky survey.

Finally, we have also generated a set of $\Lambda$CDM mock catalogues of 6dFGSv, based on the GiggleZ $N$-body simulation and incorporating the 6dFGSv selection function, to be used for testing systematic biases in the dataset. We find the 6dFGSv bulk flow amplitude is consistent with the distribution measured in the mocks. Using the mocks, we also estimate the additional uncertainty in our bulk flow amplitude, due to cosmic variance in the Fundamental Plane zero-point, to be $\sim50\kms$.

These mocks are available on request for further analyses of the 6dFGSv sample. The C++ code written to calculate the Minimum Variance bulk flow for this paper is publicly available on GitHub, at https://github.com/mscrim/MVBulkFlow.

%***********************************************************************************************************************************************
\section{Acknowledgements}

We thank Hume Feldman and Richard Watkins for assistance with the MV implementation, and we also thank Hume and Pirin Erdo{\u g}du for helpful comments on this draft. Thanks also to Matt George and Jon Carrick for helpful discussions.

M.I.S. acknowledges financial support from a Jean Rogerson Scholarship, a UWA Top-up Scholarship from the University of Western Australia, and a CSIRO Malcolm McIntosh Lecture bankmecu scholarship. M.I.S. thanks the Astronomical Society of Australia for providing financial support via a Student Travel Award, which enabled furthered collaboration on this paper, and also Lawrence Berkeley National Laboratory for hosting her during part of this work.  CB and TMD acknowledge the support of the Australian Research Council through the award of Future Fellowships, grants FT110100639 and  FT100100595 respectively.
 The Centre for All-sky Astrophysics is an Australian Research Council Centre of Excellence, funded by grant CE110001020. 

\bibliographystyle{hapj}

\newcommand{\nat}{Nat}
\newcommand{\mnras}{MNRAS}
\newcommand{\apj}{ApJ}
\newcommand{\apjl}{ApJL}
\newcommand{\apjs}{ApJS}
\newcommand{\aap}{A \& A}
\newcommand{\aj}{AJ}
\newcommand{\pasa}{PASA}
\newcommand{\prd}{Phys. Rev. D}
\newcommand{\physrep}{Phys. Rep.}
\newcommand{\na}{NewA}
\newcommand{\jcap}{JCAP}
\newcommand{\inprep}{in preparation}
\newcommand{\submitted}{submitted}

\bibliography{6dFGS_bulkflow}

%***********************************************************************************************************************************************
\appendix

%***********************************************************************************************************************************************
\section{Minimum Variance Bulk Flow Method from Watkins et al. (2009)}
\label{MV_appendix}

For a dataset consisting of $N$ peculiar velocities with positions $\textbf{\textit{r}}_n = x_i$, where $i$ indicates the 3 directions $(x,y,z)$, and measured radial peculiar velocities $S_n$, the MV method \citep{watkins2009,feldman2010} constructs a set of weights $w_{i,n}$ such that the bulk flow is given by equation~\ref{eqn_BF_estimate}. The weights act to minimise the variance between the bulk flow moments measured by the survey, $u_i$, and the bulk flow moments that would be measured by an `ideal' survey, $U_i$.

To calculate the weights, the authors apply constraints to ensure that the estimator gives the correct average amplitudes for the velocity moments, i.e. $\langle u_i \rangle = U_i$, of the form
\begin{equation}
\label{eqn_bfconstraints}
%\sum_n w_{i,n} \hat{r}_{j,n} = \delta_{ij}.
\sum_n w_{i,n} g_j({\bf r}_n) = \delta_{ij}.
\end{equation}
Here, $g_j({\bf r})$ are the mode functions corresponding to given moments of the velocity field; for the three bulk flow moments, they are
\begin{equation}
g_j({\bf r}) = \{ \hat{\bf r}_x, \hat{\bf r}_y, \hat{\bf r}_z \}.
\end{equation}
The authors implement the set of constraints in Equation~\ref{eqn_bfconstraints} using Lagrange multipliers, and so the quantity to be minimised is
\begin{equation}
\langle (U_i - u_i)^2 \rangle + \sum_j \lambda_{ij} \left[ \sum_n w_{i,n} g_j({\bf r}_n) - \delta_{ij} \right].
\end{equation}

\cite{feldman2010} show that the weights can be evaluated as:
\begin{equation}
w_{i,n} = \sum_m G_{nm}^{-1} \left( Q_{im} - {1 \over 2} \sum_j \lambda_{ij}g_j({\bf r}_m) \right).
\end{equation}
We define the matrices $G$, $Q$ and $\lambda$ below.

\subsection{Velocity covariance matrix, G}

$G_{nm} = \langle S_nS_m \rangle$ is the covariance matrix for the individual velocities, which can be calculated for a given power spectrum. In linear theory it can be written in terms of the velocity field $\textbf{\textit{v}}(\textbf{\textit{r}})$ as
\begin{eqnarray}
\label{eqn_Gnm}
G_{nm} &=& \langle S_n S_m \rangle \nonumber \\
&=& \langle v_n v_m \rangle + \delta_{nm} (\sigma_*^2 + \sigma_n^2). 
%&=&   \frac{\Omega_{\rm m}^{1.1} H_0^2}{2 \pi^2} \int {\rm d} k \, P(k) f_{mn}(k) + \delta_{nm} (\sigma_*^2 + \sigma_n^2)
\end{eqnarray} 
The first, `geometrical' term can be expressed as an integral over the density power spectrum $P(k)$:
\begin{equation}
\langle v_n v_m \rangle = \frac{f(\Omega_{\rm m},z)^{2} H_0^2 a^2}{2 \pi^2} \int {\rm d} k \, P(k) f_{mn}(k),
\end{equation}
where $H_0$ is the Hubble constant in units of $(h\textrm{ km s}^{-1}\textrm{Mpc}^{-1})$, $a$ is the cosmological scale factor, essentially equal to unity for the low redshifts we are considering, and the function $f_{mn}(k)$ is the angle averaged window function,
\begin{equation}
\label{eqn_fmnk}
f_{mn}(k) = \int \frac{{\rm d}^2\hat{k}}{4 \pi} (\hat{\textbf{\textit{r}}}_n \cdot \hat{\textbf{\textit{k}} })(  \hat{\textbf{\textit{r}}}_m \cdot  \hat{\textbf{\textit{k}}} ) \times \exp [ik\hat{\textbf{\textit{k}}} \cdot ( \textbf{\textit{r}}_n-\textbf{\textit{r}}_m)].
\end{equation}
This equation can be calculated analytically, as shown in the appendix of \cite{ma2011}.

\subsection{Velocity-bulk flow cross correlation, $Q$}

The correlation matrix $Q_{i,n}$ is calculated in a similar way, but incorporates the window function of the input `ideal' survey. It is evaluated by generating an ideal survey with $N'$ random positions $\bf{r}'_{n'}$ with the desired radial distribution function. $Q_{i,n}$ is then given by
\begin{equation}
Q_{i,n} = \langle U_iv_n \rangle = \sum_{n'=1}^{N'} w'_{i,n'} \langle v_{n'}v_n \rangle .
\end{equation}
The weights $w'_{i,n'}$ for the ideal survey simply give the bulk flow as the average of the projections of the radial velocities on the three coordinate axis directions,
\begin{equation}
w_{i,n} = \frac{3\hat{\vc{x}}_i \cdot \hat{\vc{r}}_n }{N}.
\end{equation}
(Note in WFH09 the factor of 3 has been omitted from this equation). 
Following \cite{watkins2009} we create an `ideal' survey with $N' = 10^4$ and a Gaussian radial density $n(r) \propto \exp(-r^2/2R_I^2)$, where $R_I$ is the effective radius of the Gaussian.

Then, we evaluate $\langle v_{n'}v_n \rangle$ by
\begin{equation}
\langle v_{n'}v_n \rangle =  {f(\Omega_{\rm m},z)^{2} H_0^2 a^2 \over 2 \pi ^2} \int dk P(k) f_{n'n}(k).
\end{equation}

\subsection{Lagrange multiplier, $\lambda$}

%where $G$ is the covariance matrix of the individual measured velocities, $G_{nm} \equiv \langle S_n S_m \rangle$. 
The Lagrange multiplier $\lambda_{ij}$ is given by
\begin{equation}
\lambda_{ij} = \sum_{l=1}^3 \left[ M_{il}^{-1} \left( \sum_{m,n} G_{nm}^{-1} Q_{lm} g_j (\textbf{r}_n) - \delta _{lj} \right) \right]
\end{equation}
where
\begin{equation}
M_{ij} = {1 \over 2} \sum_{n,m} G^{-1}_{nm} g_i(\textbf{r}_n) g_j(\textbf{r}_m).
\end{equation}
For the bulk flow, with $g_i(\textbf{r}) = \hat{r}_i$, the latter equation becomes
\begin{equation}
M_{ij} = {1 \over 2} \sum_{n,m} G^{-1}_{nm} \hat{r}_i(n) \hat{r}_j(m).
\end{equation}

\end{document}